\documentclass[twocolumn,aps,prd,epsf,showpacs]{revtex4-1}
\usepackage[pdftex]{graphicx}
\usepackage{epsfig}
\usepackage{dcolumn}
\usepackage{color}
\usepackage{enumitem}
\usepackage[toc,page]{appendix}
\usepackage{float}

\graphicspath{ {Figuras/} }

\begin{document}

\title{QCD vacuum replicas are metastable}
\author{P. Bicudo, J.E. Ribeiro}
\affiliation{Dep. F\'{\i}sica and CeFEMA, Instituto Superior T\'ecnico,
Av. Rovisco Pais, 1049-001 Lisboa, Portugal}
\begin{abstract}
A metastable phase has important physical implications, since it may form vacuum bubbles detectable experimentally,
whereas an unstable vacuum instantly explodes.
It is well known, due to chiral symmetry breaking, there are at least two very different QCD vacua.
At T=0, and in the true vacuum, the scalar and pseudo-scalar, 
or the vector and axial vector are not degenerate, and in the chiral limit, $m_{\left\{ q,\bar{q}\right\}}=0$,
the pseudoscalar ground states are Goldstone bosons.
At T=0, the chiral invariant vacuum is unstable, decaying through 
an infinite number of scalar and pseudo-scalar tachyons.  
Moreover, QCD vacuum replicas, an infinite tower of finite volume, excited vacuum-like solutions, with energy densities lying between the true vacuum and the chiral invariant vacuum energy densities, have been predicted due to the non-linearity of the mass gap equation with a confining interaction.
It remained to show whether the QCD replicas are metastable or unstable.
In this paper, the spectrum of quark-antiquark systems is studied both in the true vacuum and in the two first excited QCD replicas. 
The mass gap equation for the vacua and the Salpeter-RPA 
equation for the mesons are solved for a simple chiral invariant and 
confining quark model approximating QCD in the Coulomb gauge.
We find no tachyons, thus showing the QCD replicas in our approach is indeed metastable.
\end{abstract}

\maketitle

\twocolumngrid

%
%
\section{Introduction}

As it is well known, in the chiral limit, the  $SU(2)_L\times SU(2)_R)$ symmetry of the QCD Lagrangian is
broken. This entails the following properties:
\begin{enumerate}[label=(\roman*)]
\item the  conspicuous absence of low lying parity multiplets for hadrons, 
\item the very small pion mass as compared, for instance, with the $\rho$ mass and,
\item the progressive restoration of parity symmetry for angular excited hadronic states. 
\end{enumerate}

In the absence of first principles, fully controllable QCD calculation of hadronic states, one must resort to effective models that satisfy the above mentioned properties. This is the case for the Gaussian approximation to QCD  \cite{Dosch:1988ha,Nefediev:2003mx,Kalashnikova:2005tr}, 
\begin{eqnarray}
&&H=\int\, d^3x \left[ \psi^{\dag}( x) \;(m_0\beta -i{\vec{\alpha}
\cdot \vec{\nabla}} )\;\psi( x)\;+
{ 1\over 2} \int d^4y\, \
\right.
\nonumber \\
&&
\overline{\psi}( x)
\gamma^\mu{\lambda^a \over 2}\psi ( x)  
g^2 \langle A_\mu^a(x) A_\nu^b(y) \rangle
\;\overline{\psi}( y)
\gamma^\nu{\lambda^b \over 2}
 \psi( y)  \ + \ \cdots\nonumber\\
 &&
\label{hamilt}
\end{eqnarray}
with the quark kernel, $K^{ab}_{\mu\nu}(x,y)$,  being defined as
\begin{equation}
K^{ab}_{\mu\nu}(x,y)=g^2 \langle A_\mu^a(x) A_\nu^b(y) \rangle.
\end{equation}
$m_0$ is the current mass of the quark. In the remainder of this paper we will set $m_0$ to zero.

There is a large literature on the form of $K^{ab}_{\mu\nu}(x,y)$ as derived from gluon configurations (see Ref. \cite{Bicudo:1998bz,Szczepaniak:2001rg} for some examples). Here we are not concerned with the actual derivation of the quark kernel in terms of gluon configurations provided it can be represented in 
Eq.(\ref{hamilt}) as,
\begin{equation}\label{potenciais}
K^{ab}_{\mu\nu}(x,y)=\delta^{ab} \Gamma_{\mu \nu} K_0^{\alpha+1}|\vec{x}-\vec{y}|^\alpha.
\end{equation}
See \cite{Bicudo:1989sh,Bicudo:1989si,Bicudo:1989sj,Bicudo:2003cy,Bicudo:2010qp} and references therein.

From all these studies, it emerged that although numerical values may vary somewhat, as for instance when one goes from harmonic to linear confining kernels, the global overall physical  picture does not. So in this paper and for numerical simplicity, we will use the harmonic confining kernel.

The model of Eq.(\ref{hamilt}) was suggested in the mid-eighties \cite{Amer:1983qa,LeYaouanc:1984ntu,LeYaouanc:1983huv}, and latter re-derived in terms of coherent states of $^3P_0$ quark-antiquark pairs \cite{Bicudo:1989sh,Bicudo:1989si,Bicudo:1989sj}.

  As it is also evident from Eq.(\ref{hamilt}), chiral symmetry is explicitly broken by quark masses, so that we are led to expect that for sufficiently heavy quark masses, a quasi classical regime for hadrons must set in in contrast to the quantum nature of low lying hadrons, the latter being entirely due to quantum fermion loops \cite{Glozman:2005tq}. Finally, it can be shown that for soft chiral symmetry breaking due to small quark masses, the Hamiltonian of Eq.(\ref{hamilt}) obey the requirements of PCAC: the Gell-Mann Oakes and Renner relation, the Adler Zero and the Weinberg Theorem \cite{LeYaouanc:1984ntu,Bicudo:1998mc,Bicudo:2001jq,
LlanesEstrada:2003ha,Bicudo:2003fp,Bicudo:2001zu,Nefediev:2006bm},

%
\begin{figure}[t]
\includegraphics[trim=60pt 60pt 0 20pt ,width=1.15\columnwidth]{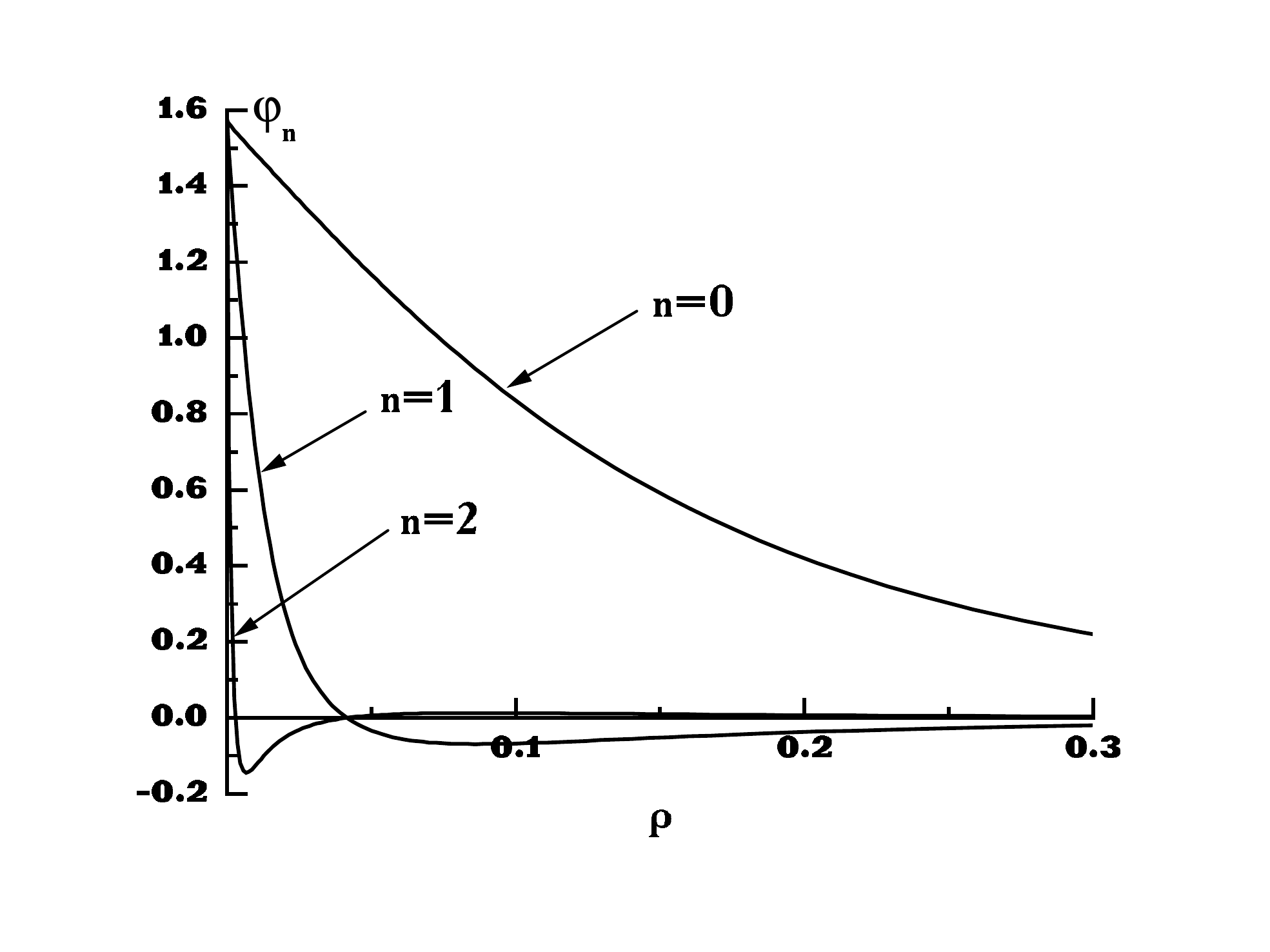}
\caption{
The first solution, $\varphi_0(k)$, of the mass gap equation, representing the vacuum $| 0 \rangle$ and its two first two other solution, $\varphi(k)_1$ and $\varphi(k)_2$ for a linear confining potential \cite{Bicudo:2003cy,Bicudo:2010qp}, corresponding to the truncated Coulomb gauge of QCD), computed in the chiral limit of $m_{q\,\;bar{q}}=0$.
}\label{fig:replicaslinear}
\end{figure} 

From the onset of chiral symmetry breaking, it is well known that there should be at least two very different classes of QCD "vacua" (defined as stationary solutions for the mass-gap equation): a chirally symmetric class of states and another, which breaks chiral invariance. This corresponds to a vacua manifold having the usual Mexican hat form, with one of the solutions being the chiral invariant state.  In the simplest scenario, as in the flavour SU(2) sigma model, and for such a chirally symmetric state, we get a finite number of tachyons, both in the scalar $\sigma$ and the pseudoscalar $\left\{ \pi^+, \ \pi^0 , \ \pi^-\right\}$ channels, amounting to four tachyons in total, with masses $M^2<0$, whereas for the chiral symmetry breaking vacuum (the true vacuum), we find one massive meson, the scalar $\sigma$, with $M^2> 0$, plus three pseudoscalar Goldstone mesons $ \left\{\pi^+, \ \pi^0 , \ \pi^-\right\}$ and no tachyons. 

However, the QCD vacuum structure is expected to be richer than the Mexican hat picture shows -- see for instance, Refs. \cite{Osipov:2002wj,Martin:2006qd}.
When we have an Hamiltonian as it is given in Eq.(\ref{hamilt}), endowed  with a quark confining potential like those of Eq.(\ref{potenciais}), the excitations of QCD are much richer. The number of tachyon states for the false chiral invariant vacuum becomes infinite 
\cite{Bicudo:2006dn},
whereas, from the true vacuum, we can build an infinite set of true hadronic states. Furthermore, it was found that for those quark kernels, there will be also a separate infinite tower of excited vacua-like states interpolating between the true chiral symmetry breaking, physical vacuum, to the highest chiral invariant vacuum \cite{LeYaouanc:1984ntu,Bicudo:1989sh,Bicudo:2003cy}. 
Examples of solutions in the case of the linear potential \cite{Bicudo:2003cy} for the chiral angle $\varphi$ 
are depicted in Fig.(\ref{fig:replicaslinear}). 

For infinite volumes, they correspond to orthogonal $^3P_0$ coherent states so that, two different coherent states, pertaining to two different chiral angle functions, have a zero overlap between them, hence  being independent of each other. In addition, they are separated apart by an infinite amount of energy. 

However, this independence sets in very fast even for relatively small volumes of the overlap. Throughout the remainder of this paper we will call replica ${\cal R}i_{[V]}$ to a $^3P_0$ coherent space corresponding to a mass-gap solution $\varphi_i(p)$, occupying a finite volume V, provided its overlap with the physical vacuum should be as close to zero so as to effectively prevent its decay. The true physical vacuum corresponds to ${\cal R}0_{[\infty]}$. As usual, we set the energy scale origin $E_{{\cal R}0}[\infty]$ to zero that is, its energy density, ${\cal E}_{{\cal R}0}\equiv 0$. When the volume goes to infinity, this replica will go to the corresponding excited vacuum and its energy, $E_{{\cal R}i}[\infty]$, will go to infinity. For finite volumes this energy is finite. From now on, to the word replica we will always mean a $3P0$ coherent state occupying a finite volume and, for this reason, we will drop $[V]$ from the notations.

\emph{ In this paper we are not concerned with the creation mechanism of ${\cal R}i$ -- see \cite{AntNefRib-2010} for a possible mechanism. Here we will show that, if and once they are created, they do not possess tachyons for any finite volume  of ${\cal R}i$, no matter how large the volume might be, and they become, already starting with a small volume and upward, to be increasingly independent of each other and, of course, of the vacuum. All the hadronic-like excitations we might build upon any of the ${\cal R}i$, will have an excess mass ${\cal M}^{{\cal P}}_i$ in relation to $E_{{\cal R}i}$ where ${\cal P}$ denotes the particular excitation we might consider and i, the particular replica ${\cal R}i$ we choose to work with. ${\cal M}^{{\cal P}}_i$ depends on the replica i, through the chiral angle $\varphi_i$. {\bf But} ${\cal M}^{{\cal P}}_i$ {\bf does not depend on $[V]$}. Of course, once a small, sufficiently stable replica is formed, the hadronic excitations therein will have a mass given by $M^{{\cal P}}_i={\cal M}^{{\cal P}}_i+E_{{\cal R}i}$. }

\begin{figure}[t!]
\includegraphics[width=0.9\columnwidth]{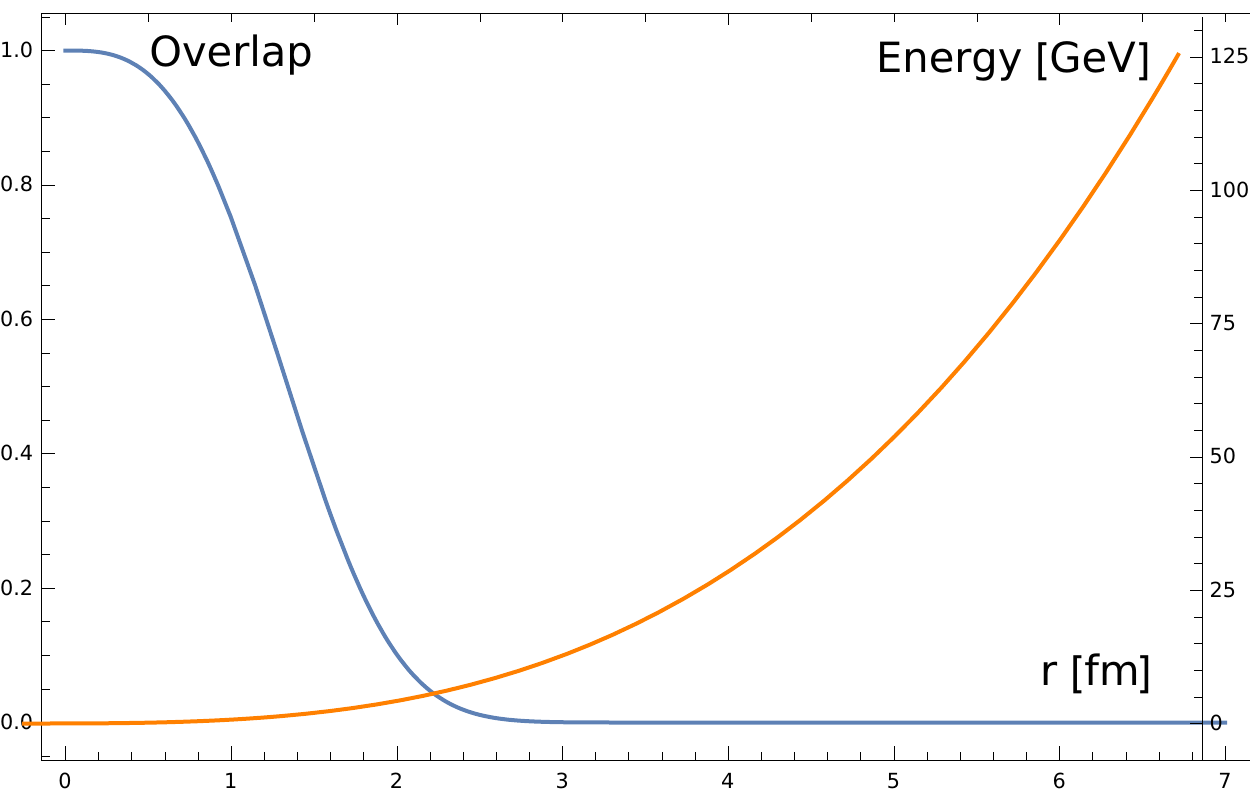}
\caption{We show (blue) the overlap $\langle {\cal R}1| 0 \rangle$ and (orange) the corresponding energy  $ E_{{\cal R}1}$ of the first replica for different spherical bubble radii. The energy density is $({\cal E}_{{\cal R}1}\simeq  87.2476\;MeV/fm^3$ (using $K_0\simeq 300 MeV$) and, for instance, with a radius of only $5$ fm, we would had to add to any hadronic excitation in ${\cal R}1$ another $45$ GeV, and this for an overlap $\langle {\cal R}1_| 0 \rangle =2.9 \times 10^{-16}$. 
In practice, due to the exponential overlap of $\langle {\cal R}1_| 0 \rangle$ (Eq.(\ref{eq:1})), we do not need an infinite volume $V$ to get a near orthogonality between the replica $| {\cal R}1\rangle$ and the vacuum $| 0 \rangle$. We do not plot the overlap for $\langle0|{\cal R}2\rangle$ because it has to be smaller than $\langle0|{\cal R}1\rangle$ -- a trivial consequence of Eq.(\ref{eq:1}).}\label{fig:ortener}
\end{figure}

As an exemplification of the above, we will consider the first two replicas and use for them a volume V, corresponding to a bubble of 5 fm radius.  Already for this volume, their  overlap with the vacuum are exceedingly small -- see Eq.(\ref{eq:1}) and Fig.(\ref{fig:ortener}). Of course, for larger volumes we would get an even smaller overlap and an increased energy cost to create $ E_{{\cal R}\left\{1,2\right\}}$. 

Then, it should be clear that the actual size of the volume does not matter for the paper conclusions, as we can always think of a succession of replicas occupying ever larger finite volumes. For each member of this set, provided that its overlap with the vacuum is sufficiently small as to ensure its stability,  there will be no tachyons. This indicates that QCD may still have a very rich structure (and one that is free of tachyonic sates) with far reaching implications \cite{Bicudo:2002eu,Nefediev:2002nw}.  

The paper is organized as follows. 

In Section II, the quark mass gap equation and the bound state 
quark-antiquark equation are reviewed. 
In Section III, the mass gap and bound state equations are solved 
numerically and we show the results for the Salpeter wave functions. The mesonic  meson spectra for the stable vacuum,  ${\cal R}1$ and  ${\cal R}2$ are also shown. The conclusion and outlook are presented in Section V.

%

%
%
\section{Mass gap and bound state equations}

\subsection{Mass gap equation}

After Wick contractions, the Hamiltonian of Eq.(\ref{hamilt}) can be formally be written as,
\begin{equation}\label{eq:0}
H =[\int d^3x]\; {\cal E}_{[\varphi]} + :H_2[\varphi]: + :H_4[\varphi]:\; ,
\end{equation}
  
In the non condensed vacuum, $\varphi$ is equal to $\arctan{m_0 / k}$,
but $\varphi$ is not determined from the onset when chiral symmetry breaking occurs. In the physical vacuum, the constituent quark mass $m_c(k)$, or the
chiral angle, $\varphi(k)=\arctan [m_c(k) / k]$, is a variational function
which should be determined by the mass gap equation. 

${\cal E}_{[\varphi]}$ is a functional of the chiral angle $\varphi(k)$ \cite{Amer:1983qa,Bicudo:1989sh,Bicudo:2010qp}, and represents the energy density corresponding to the non-perturbative vacuum condensation with $^3P_0$ quark-antiquark pairs. It negative when compared with the trivial false vacuum of $\varphi=0$, and is given by 
\cite{Bicudo:1989sh,Bicudo:2010qp},
\begin{eqnarray}\label{conEner}
&&{\cal E}_{[\varphi]}=- {N_s N_cN_f \over 2} \int {d^3 k \over (2 \pi)^3} \left[ E(k)+m_0\, S(k)+ k \, C(k) \right]
\nonumber \\
&&E_q(k)=A(k) S(k)+B(k) C(k) 
\nonumber\\
&&A(p)=m_0 + \frac{1}{2} \int {d^3 k \over (2 \pi)^3} V (\vec{p}-\vec{k}) S(k)
\nonumber\\
&&B(p)=p + \frac{1}{2} \int {d^3 k \over (2 \pi)^3} V (\vec{p}-\vec{k}) \left(\hat{p}\cdot\hat{k}\right) C(k)
\end{eqnarray}
with $E_q(k)$ denoting the quark energy. Finally, $:H_2[\varphi]:$ and $:H_4[\varphi]:$ stand for the quadratic and quartic fermionic terms for the same $\varphi(k)$.

%
%
\begin{table}[t!]
\begin{ruledtabular}
\begin{tabular}{c|ccccc}
$^{2S+1}L_J$					& $\delta_{{\bf S}_q,{\bf S}_{\bar q}}$ 
									& $ {\bf S}_q \hspace{-.075 cm} 
									\cdot \hspace{-.075 cm} {\bf S}_{\bar q}$ 
											& $({\bf S}_q + {\bf S}_{\bar q}) \hspace{-.075 cm} 
											\cdot \hspace{-.075 cm} {\bf L}$
													& $({\bf S}_q - {\bf S}_{\bar q})\hspace{-.075 cm} 
													\cdot \hspace{-.075 cm} {\bf L}$
														& tensor		\\
$^1S_0$ 						&1	&-3/4	&0		&0	&0 				\\
$^3P_0$ 						&1	&1/4	&-2		&0	&-1/3 			\\
$^3S_1$ 						&1	&1/4	&0		&0	&0 				\\
$^3D_1$ 						&1	&1/4	&-3		&0	&-1/6 			\\
$^3S_1\leftrightarrow {}^3D_1$	&0	&0		&0		&0	&$\sqrt{2}$/6 	\\
$^1P_1$ 						&1	&-3/4	&0		&0	&0 				\\
$^3P_1$ 						&1	&1/4	&-1		&0	&1/6 			\\
$^1P_1\leftrightarrow {}^3P_1$	&0	&0		&0		&$\sqrt{2}$	&0			 	\\
\end{tabular}
\end{ruledtabular}
\caption{
Matrix elements of the spin-dependent potentials
}
\label{algebraic} 
\end{table}

The relativistic invariant Dirac-Feynman propagators
\cite{LeYaouanc:1984ntu,Bicudo:1998mc}, 
can be decomposed in the quark and antiquark Bethe-Goldstone 
propagators, close to the formalism of non-relativistic quark models,
\begin{eqnarray}
&&{\cal S}_{Dirac}(k_0,\vec{k})
={i \over \not k -m +i \epsilon}=\nonumber\\
&&\quad\quad= {\sum_su_su^{\dagger}_s \beta \over k_0 -E(k) +i \epsilon} 
- {\sum_sv_sv^{\dagger}_s \beta  \over -k_0 -E(k) +i \epsilon}
\end{eqnarray}
with,
\begin{eqnarray}
u_s({\bf k})&=& \left[
\sqrt{ 1+S \over 2} + \sqrt{1-S \over 2} \widehat k \cdot \vec \sigma \gamma_5
\right]u_s(0)  \ ,
\nonumber \\
v_s({\bf k})&=& \left[
\sqrt{ 1+S \over 2} - \sqrt{1-S \over 2} \widehat k \cdot \vec \sigma \gamma_5
\right]v_s(0)  \ ,
\nonumber \\
&=& -i \sigma_2 \gamma_5 u_s^*({\bf k}) \ ,
\label{propagators}
\end{eqnarray}
where we use the simplified notation $S=\sin(\varphi(k))={m_c(k)\over \sqrt{k^2+m_c(k)^2}}$ and
$\ C=\cos(\varphi(k))={k\over \sqrt{k^2+m_c(k)^2}}$.

\subsection{Replicas}
Central to the Hamiltonian of Eq.(\ref{hamilt}) we have the so called mass gap.  Being inherently non linear, it supports more than one solution. Let us denote such solutions by the functions $\varphi_i(k)$, It turns out that these solutions can be given in terms of corresponding wave function for the $^3P_0$ quark-antiquark coherent state  ${\cal R}i$ \cite{Bicudo:1989sh}. Furthermore, for each and every two such solutions, the overlap between the corresponding $^3P_0$ quark-antiquark coherent states will behave like \cite{Nefediev:2002nw},
\begin{equation}
\langle {\cal R}i|{\cal R}j\rangle =\exp\left\{
    V N_c N_F\int \frac{d^3p}{(2\pi)^3}  \log
    \left\{\cos^2\left[\frac{\delta \varphi (p)}{2} \right]\right\}
    \right\},
\label{eq:1}
\end{equation}
with $\delta \varphi (p)=\varphi(p)_{{\cal R}i}-\varphi(p)_{{\cal R}j}$. $N_c$,$N_F$ stand respectively for the number of colours and flavours and $V$ is the spacial volume. We have, $\lim_{V\to \infty} \langle {\cal R}i|{\cal R}j\rangle_{[V]} = 0  $ for any different  $\varphi_i$ and $\varphi_j$. This means that in the infinite volume, $[V]=\infty$, the Fock space, ${\cal F}_{{\cal R}i}[\infty]$ is orthogonal to  ${\cal F}_{{\cal R}j}[\infty]$  and no fermion in 
${\cal F}_{{\cal R}j}[\infty]$ admits a Fourier decomposition in terms of operators in ${\cal F}_{{\cal  R}i}[\infty]$. Among all the $^3P_0$ quark-antiquark coherent states $|{\cal R}i[\infty]\rangle$, the one with the lowest energy density ${\cal E}_{[\varphi_0]}$, is called the true vacuum $|{\cal R} 0_{[\infty]}\equiv|0\rangle$. 

As we said, for the true vacuum, the energy density, ${\cal E}_0\equiv {\cal E}_{{\cal R}0}$ is set as the origin of the energies and, therefore, is discarded. Notice that once this is done, the same cannot be done for any of the remaining  energy densities ${\cal E}_{{\cal R}i}$. In the remainder of this paper we will focus on the true vacuum and on the first two replicas, ${\cal R}i,\;i=\left\{1,2\right\}$.

We get for their energy densities,
\begin{eqnarray}
{\cal E}_{{\cal R}1}= 0.0861 {K_0}^4 = 87.2 \text{MeV / fm}^3 \ ,
\nonumber \\
{\cal E}_{{\cal R}2}=0.0941 {K_0}^4 = 95.3 \text{MeV / fm}^3 \ ,
\end{eqnarray}
for a $K_0\simeq 300MeV$. For the exponent of the overlap $\langle {\cal R}1| 0 \rangle$ in Eq.(\ref{eq:1}) we get $- V \times 0.00759$ fm$^{-}3$.
In Fig.(\ref{fig:ortener}), we depict the replica-vacuum overlap $\langle {\cal R}1| 0 \rangle$ together with its energy  $E_{{\cal R}1}$, in terms of the radius of  a spherical replica.

\subsection{Mass gap}
There are three equivalent methods to solve the mass gap equation:
\begin{enumerate}[label=(\roman*)]
\item assume a quark-antiquark $^3P_0$ 
coherent state, and minimize its correspondent energy density,
\item
rotate the quark fields by a Bogoliubov-Valatin canonical transformation to diagonalize the Hamiltonian,
\item
solve the Schwinger-Dyson equations for the quark propagators. 
\end{enumerate}

%
%
\begin{table}[t!]
\begin{ruledtabular}
\begin{tabular}{c|c}
& $V^{++}=V^{--}$  \\ \hline
spin-indep. & $- {d^2 \over dk^2 } + { {\bf L}^2 \over k^2 } + 
{1 \over 4} \left( {\varphi'_q}^2 + {\varphi'_{\bar q}}^2 \right) 
+ {  1 \over k^2} \left( {\cal G}_q +{\cal G}_{\bar q}  \right) $  \\ 
spin-spin & $ {4 \over 3 k^2} {\cal G}_q {\cal G}_{\bar q} {\bf S}_q \cdot {\bf S}_{\bar q} $  \\ 
spin-orbit & $ {1 \over  k^2} \left[ \left( {\cal G}_q + 
{\cal G}_{\bar q} \right) \left( {\bf S}_q +{\bf S}_{\bar q}\right) 
+\left( {\cal G}_q - {\cal G}_{\bar q} \right) \left( {\bf S}_q -{\bf S}_{\bar q}\right)  \right]
\cdot {\bf L} $  \\ 
tensor & $ -{2 \over  k^2} {\cal G}_q {\cal G}_{\bar q} 
\left[ ({\bf S}_q \cdot \hat k ) ({\bf S}_{\bar q} \cdot \hat k )
-{1 \over 3} {\bf S}_q \cdot {\bf S}_{\bar q} \right] $ \\ \hline
& $V^{+-}=V^{-+}$ \\ \hline
spin-indep. & $0$  \\ 
spin-spin & $ -{4 \over 3} \left[ {1\over 2} {\varphi'_q} {\varphi'_{\bar q}} + 
{1\over k^2} {\cal C}_q {\cal C}_{\bar q}  \right]
{\bf S}_q \cdot {\bf S}_{\bar q} $  \\ 
spin-orbit & $0$  \\ 
tensor & $ \left[ -2 {\varphi'_q} {\varphi'_{\bar q}} + 
{2\over k^2} {\cal C}_q {\cal C}_{\bar q}  \right]
\left[ ({\bf S}_q \cdot \hat k ) ({\bf S}_{\bar q} \cdot \hat k )
-{1 \over 3} {\bf S}_q \cdot {\bf S}_{\bar q} \right] $
\end{tabular}
\end{ruledtabular}
\caption{
The positive and negative energy spin-independent, spin-spin, spin-orbit and 
tensor potentials are shown, for the simple harmonic model \cite{LeYaouanc:1984ntu}.
$\varphi'(k)$, ${\cal C}(k)$ and ${\cal G}(k)= 1 - S(k) $ are all functions of the constituent 
quark(antiquark) mass.
}
\label{spindependent} 
\end{table}

Any of these methods
lead to the same mass gap equation and $m_c(k)$. Here, we replace the propagator
of Eq. (\ref{propagators}) in the Schwinger-Dyson equation, 
{\small \begin{eqnarray}
\label{2 eqs}
&&0 = u_s^\dagger(k) \left\{k \widehat k \cdot \vec \alpha + m_0 \beta
-\int {d w' \over 2 \pi} {d^3k' \over (2\pi)^3}
i V(k-k') \right.
\nonumber \\
&&\left. \sum_{s'} \left[ { u(k')_{s'}u^{\dagger}(k')_{s'} 
 \over w'-E(k') +i\epsilon}
-{ v(k')_{s'}v^{\dagger}(k')_{s'} 
  \over -w'-E(k')+i\epsilon} \right]
\right\} v_{s''}(k) \  \
\nonumber \\
&&E(k) = u_s^\dagger(k) \left\{k \widehat k \cdot \vec \alpha + m_0 \beta
-\int {d w' \over 2 \pi} {d^3k' \over (2\pi)^3}
i V(k-k')  \right.
\nonumber \\
&&\left. \sum_{s'} \left[ { u(k')_{s'}u^{\dagger}(k')_{s'} 
 \over w'-E(k') +i\epsilon}
-{   v(k')_{s'}v^{\dagger}(k')_{s'}  
 \over -w'-E(k')+i\epsilon} \right]
\right\} u_s(k),
\end{eqnarray}        }

It was shown that for the wide class of confining potentials \cite{Bicudo:2003cy}, with the exponent $ 0<\alpha \leq 2 $ -- see Eq.(\ref{potenciais}) -- it leads to an infinite number of solutions of the mass gap equation. The two cases mostly studied in the literature are the quadratic case, $\alpha=2$, derived in the Gaussian approximation to QCD and using the Balitsky Local Coordinate Gauge \cite{Bicudo:1998bz}, and the linear case $\alpha=1$, derived in the Coulomb gauge. 
In the case of a linear confining potential, we get integral equations. They are finite but need a regularization of infrared divergences. 
Since the case of a linear confining potential is more difficult to control numerically. Here we specialize to the case of the quadratic confining potential.

With the simple harmonic interaction
\cite{LeYaouanc:1984ntu}, 
the integral of the potential is a Laplacian, and the mass gap equation can be transformed into a differential equation. The mass gap equation and the quark energy are finally given by,
{\small \begin{eqnarray}
\label{mass gap}
&&\Delta \varphi(k) = 2 k S(k) -2 m_0 C(k) - { 2 S(k) C(k) \over k^2 } \nonumber  \\ 
&&E(k) = k C(k) + m_0 S(k) - { {\varphi'(k) }^2 \over 2 } - { C(k)^2 \over k^2 }  \ .
\end{eqnarray}          }

Numerically, this equation is a non-linear ordinary differential
equation. It can be solved with the Runge-Kutta and shooting method.

\begin{figure}[t]
\includegraphics[width=0.80\columnwidth]{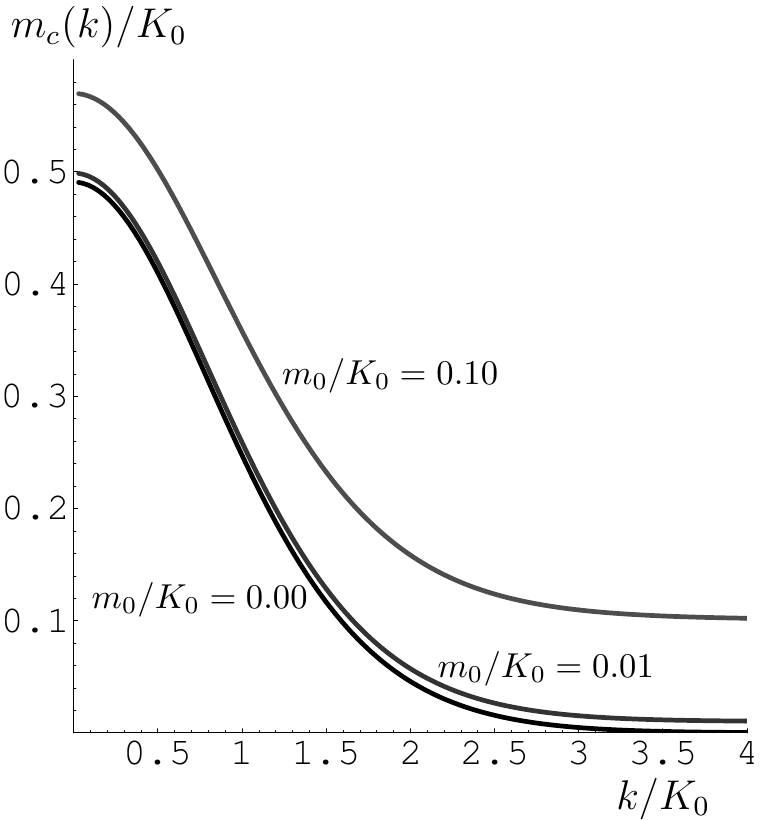}
\caption{
The dependence \cite{Bicudo:2006dn} of the constituent quark mass $m_c(k)$ on the current quark mass $m_0$.
}\label{fig:masssolution}
\end{figure}

Examples of solutions,
for different light current quark masses $m_0$, 
are depicted in Fig.(\ref{fig:masssolution}). The effect of a small finite current quark mass $m_0 \sim 0.01 K_0 $, typical of the light $u$ and $d$ quarks, can be easily estimated as a small increase of the dynamically generated constituent quark mass $m_c$ and does not concern us here.

%

\subsection{Salpeter-RPA equations for mesons}

The Salpeter-RPA equations for a meson (a colour singlet
quark-antiquark bound state) can be derived either from the $q\bar{q}$ Lippman-Schwinger
equations, or by replacing the propagator
of Eq.(\ref{propagators}) in the Bethe-Salpeter equation. In either way, one gets
\cite{Bicudo:1998mc},
\begin{eqnarray}
\chi(k,P) &=&
\int {d^3k' \over (2\pi)^3} V(k-k') \left[ 
u(k'_1)\phi^+(k',P)v^\dagger(k'_2) \right.
\nonumber \\
&&\left. +v(k'_1){\phi^-}^t(k',P) u^\dagger(k'_2)\right] 
\end{eqnarray}
together with,\vspace{-0.5cm}
\begin{eqnarray}
\label{homo sal}
\phi^+(k,P) &=& { u^\dagger(k_1) \chi(k,P)  v(k_2) 
\over +M(P)-E(k_1)-E(k_2) }
\nonumber \\
{\phi^-}^t(k,P) &=& { v^\dagger(k_1) \chi(k,P) u(k_2)
\over -M(P)-E(k_1)-E(k_2)},
\end{eqnarray}
where $k_1=k+{P \over 2} \ , \ k_2=k-{P \over 2}$ and $P$ is
the total momentum of the meson. Notice that solving for $\chi$, one gets the Salpeter equations of Le Yaouanc et al.
\cite{LeYaouanc:1984ntu}. The Salpeter-RPA equations 
\cite{Bicudo:1989sh,LlanesEstrada:1999uh}
are obtained deriving the equation for the positive 
$\phi^+$ and negative $\phi^-$ energy wave functions. The 
relativistic equal time equations have the double of coupled
equations than the Schr\"odinger equation, although in many cases, the
negative energy components can be quite small. 
The Pauli $\vec \sigma$ matrices in the spinors of Eq.(\ref{propagators}), 
produce the spin-dependent
\cite{Bicudo:1991kz} 
potentials of Table \ref{spindependent}. Notice that both the pseudoscalar and scalar equations
have a system with two equations. This is the minimal number of relativistic 
equal time equations. However, the spin-dependent interactions 
couple an extra pair of equations, both in the vector and axial-vector channels.
While the coupling of the s-wave and the d-wave are standard in vectors, the coupling 
of the spin-singlet and spin-triplet in axial-vectors only occurs if the quark and antiquark 
masses are different
${m_0}_q \neq {m_0}_{\bar q} $, say, in heavy-light systems. We now combine the algebraic matrix elements of Table \ref{algebraic}
with the spin-dependent potentials of Table \ref{spindependent},
to derive the full Salpeter-RPA radial bound state
equations. 
Using the harmonic oscillator potential \cite{Bicudo:2006dn}, 
we thus obtain the Salpeter-RPA equations for pseudoscalar, scalar, vector and axial vector mesons:

\begin{widetext}
\bigskip
\noindent {\bf the $J^{PC}=0^{-+}$,  $^1 S_0$ pseudoscalar ($P$) equations,}
\smallskip
{\small \begin{equation}
\label{pseudoscalar}
\left\{ \left( -{d^2 \over d k ^2} +E_q(k) +E_{\bar q}(k)  
+ { {\varphi'_q}^2 +{\varphi'_{\bar q}}^2 \over 4} + {1-S_q S_{ \bar q} \over k^2}  \right)
\left[ \begin{array}{cc}
1 & 0 \\ 0 & 1 \end{array} \right]
+
\left( {\varphi'_q \varphi'_{\bar q} \over 2} + {C_q C_{\bar q} \over k^2 } \right)
\left[ \begin{array}{cc}
0 & 1 \\ 1 & 0 \end{array} \right]
-{\cal M} 
\left[ \begin{array}{cc}
1 & 0 \\ 0 & -1 \end{array} \right]
\right\}
\left( \begin{array}{c} \nu_{^1S_0}^+(k) \\ \nu_{^1S_0}^-(k) \end{array} \right) = 0 
\  ;
\end{equation}   }
\\[0.5cm]
\noindent {\bf the $J^{PC}=0^{+}$, $^3 P_0$ scalar ($S$) equations,} 
\smallskip
{\small \begin{equation}
\label{scalar}
\left\{ \left( -{d^2 \over d k ^2} +E_q(k) +E_{\bar q}(k)  
+ { {\varphi'_q}^2 +{\varphi'_{\bar q}}^2 \over 4} + {1+S_q S_{ \bar q} \over k^2}  \right)
\left[ \begin{array}{cc}
1 & 0 \\ 0 & 1 \end{array} \right]
+
\left( {\varphi'_q \varphi'_{\bar q} \over 2} - {C_q C_{\bar q} \over k^2 } \right)
\left[ \begin{array}{cc}
0 & 1 \\ 1 & 0 \end{array} \right]
-{\cal M} 
\left[ \begin{array}{cc}
1 & 0 \\ 0 & -1 \end{array} \right]
\right\}
\left( \begin{array}{c} \nu_{^3P_0}^+(k) \\ \nu_{^3P_0}^-(k) \end{array} \right) = 0
\  ;
\end{equation} }
\\[0.5cm]
\noindent {\bf the $J^{PC}=1^{--}$,  coupled $^3 S_1$ and $^3D_1$ vector ($V$ and $V^*$) equations,}
\smallskip
{\small \begin{eqnarray}
\label{vector}
\left\{ 
\left( -{d^2 \over d k ^2} +E_q(k) +E_{\bar q}(k)  
+ { {\varphi'_q}^2 +{\varphi'_{\bar q}}^2 \over 4} 
+ {7-4S_q -4S_{ \bar q}+S_q S_{ \bar q}\over 3 k^2} \right)
\left[ \begin{array}{cccc}
1 & 0 & 0 & 0 \\ 
0 & 1 & 0 & 0 \\ 
0 & 0 & 0 & 0 \\ 
0 & 0 & 0 & 0 
\end{array} \right]
+
\left( -{\varphi'_q \varphi'_{\bar q} \over 6} 
-{C_q C_{\bar q} \over 3k^2 } \right)
\left[ \begin{array}{cccc}
0 & 1 & 0 & 0 \\ 
1 & 0 & 0 & 0 \\
0 & 0 & 0 & 0 \\
0 & 0 & 0 & 0
\end{array} \right]
\right. &&
\nonumber \\ 
+
\left( -{d^2 \over d k ^2} +E_q(k) +E_{\bar q}(k)  
+ { {\varphi'_q}^2 +{\varphi'_{\bar q}}^2 \over 4} 
+ {8+4S_q +4S_{ \bar q}+2S_q S_{ \bar q}\over 3 k^2} \right)
\left[ \begin{array}{cccc}
0 & 0 & 0 & 0 \\ 

0 & 0 & 0 & 0 \\ 
0 & 0 & 1 & 0 \\ 
0 & 0 & 0 & 1 
\end{array} \right]
+
\left(  {\varphi'_q \varphi'_{\bar q} \over 6} 
-{ 2 C_q C_{\bar q} \over 3k^2 } \right)
\left[ \begin{array}{cccc}
0 & 0 & 0 & 0 \\ 
0 & 0 & 0 & 0 \\
0 & 0 & 0 & 1 \\
0 & 0 & 1 & 0
\end{array} \right] 
&&
\nonumber \\ 
\left.
- 
{ \left(1-S_q\right) \left(1-S_{ \bar q}\right)\over 3 k^2 } 
\left[ \begin{array}{cccc}
0 & 0 & \sqrt{2} & 0 \\ 
0 & 0 & 0 & \sqrt{2} \\ 
\sqrt{2} & 0 & 0 & 0 \\ 
0 & \sqrt{2} & 0 & 0 

\end{array} \right]
-\left(  {\varphi'_q \varphi'_{\bar q} \over 3} 
- {C_q C_{\bar q} \over 3k^2 } \right) 
\left[ \begin{array}{cccc}
0 & 0 & 0 & \sqrt{2} \\ 
0 & 0 & \sqrt{2} & 0 \\
0 & \sqrt{2} & 0 & 0 \\
\sqrt{2} & 0 & 0 & 0
\end{array} \right]
-{\cal M} 
\left[ \begin{array}{cccc}
1 & 0 & 0 & 0 \\ 
0 & -1 & 0 & 0 \\ 
0 & 0 & 1 & 0 \\ 
0 & 0 & 0 & -1 
\end{array} \right]
\right\} 
\left( \begin{array}{c} \nu_{^3S_1}^+(k) \\ \nu_{^3S_1}^-(k) \\ \nu_{^3D_1}^+(k) \\ \nu_{^3D_1}^-(k) 
\end{array} \right) 
& =& 0
\  ; \nonumber\\
&&
\end{eqnarray}   }
\\[0.5cm]
\noindent {\bf the $J^P=1^{+}$, $^1P_1$ and $^3P_1$ axialvector ($A$ and $A^*$) equations,}
\smallskip
{\small 
\begin{eqnarray}
\label{axialvector}
\left\{ \left( -{d^2 \over d k ^2} +E_q(k) +E_{\bar q}(k)  
+ { {\varphi'_q}^2 +{\varphi'_{\bar q}}^2 \over 4} 
+ {3-S_q S_{ \bar q} \over k^2}  \right)
\left[ \begin{array}{cccc}
1 & 0 & 0 & 0 \\ 
0 & 1 & 0 & 0 \\ 
0 & 0 & 0 & 0 \\ 
0 & 0 & 0 & 0 
\end{array} \right]
+
\left( {\varphi'_q \varphi'_{\bar q} \over 2} + {C_q C_{\bar q} \over k^2 } \right)
\left[ \begin{array}{cccc}
0 & 1 & 0 & 0 \\ 
1 & 0 & 0 & 0 \\
0 & 0 & 0 & 0 \\
0 & 0 & 0 & 0
\end{array} \right]
\right.
&&
\nonumber \\ 
\left( -{d^2 \over d k ^2} +E_q(k) +E_{\bar q}(k)  
+ { {\varphi'_q}^2 +{\varphi'_{\bar q}}^2 \over 4} 
+{2 \over k^2} \right)
\left[ \begin{array}{cccc}
0 & 0 & 0 & 0 \\ 
0 & 0 & 0 & 0 \\ 
0 & 0 & 1 & 0 \\ 
0 & 0 & 0 & 1 
\end{array} \right]
+
\left( - {\varphi'_q \varphi'_{\bar q} \over 2} \right)
\left[ \begin{array}{cccc}
0 & 0 & 0 & 0 \\ 
0 & 0 & 0 & 0 \\
0 & 0 & 0 & 1 \\
0 & 0 & 1 & 0
\end{array} \right] 
&&
 \nonumber \\
\left.
+
{ S_q - S_{\bar q}\over  k^2 } 
\left[ \begin{array}{cccc}
0 & 0 & \sqrt{2} & 0 \\ 
0 & 0 & 0 & \sqrt{2} \\ 
\sqrt{2} & 0 & 0 & 0 \\ 
0 & \sqrt{2} & 0 & 0 
\end{array} \right]
-{\cal M} 
\left[ \begin{array}{cccc}
1 & 0 & 0 & 0 \\ 
0 & -1 & 0 & 0 \\ 
0 & 0 & 1 & 0 \\ 
0 & 0 & 0 & -1 
\end{array} \right]
\right\} 
\left( \begin{array}{c} \nu_{^1P_1}^+(k) \\ \nu_{^1P_1}^-(k) \\ \nu_{^3P_1}^+(k) \\ \nu_{^3P_1}^-(k) 
\end{array} \right) 
& =& 0
\  . 
\end{eqnarray}}
\end{widetext}
\pagebreak
\begin{figure}[t!]
\includegraphics[width=0.66\columnwidth]{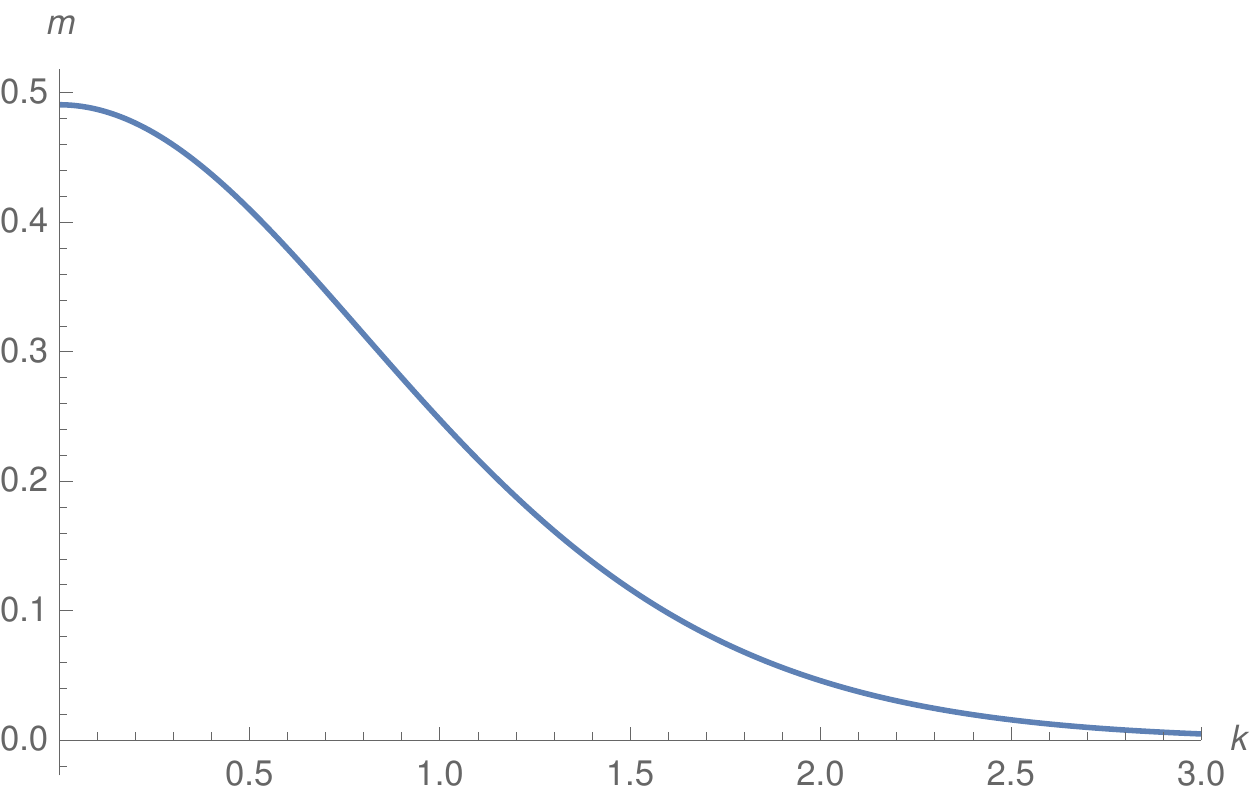}
\includegraphics[width=0.66\columnwidth]{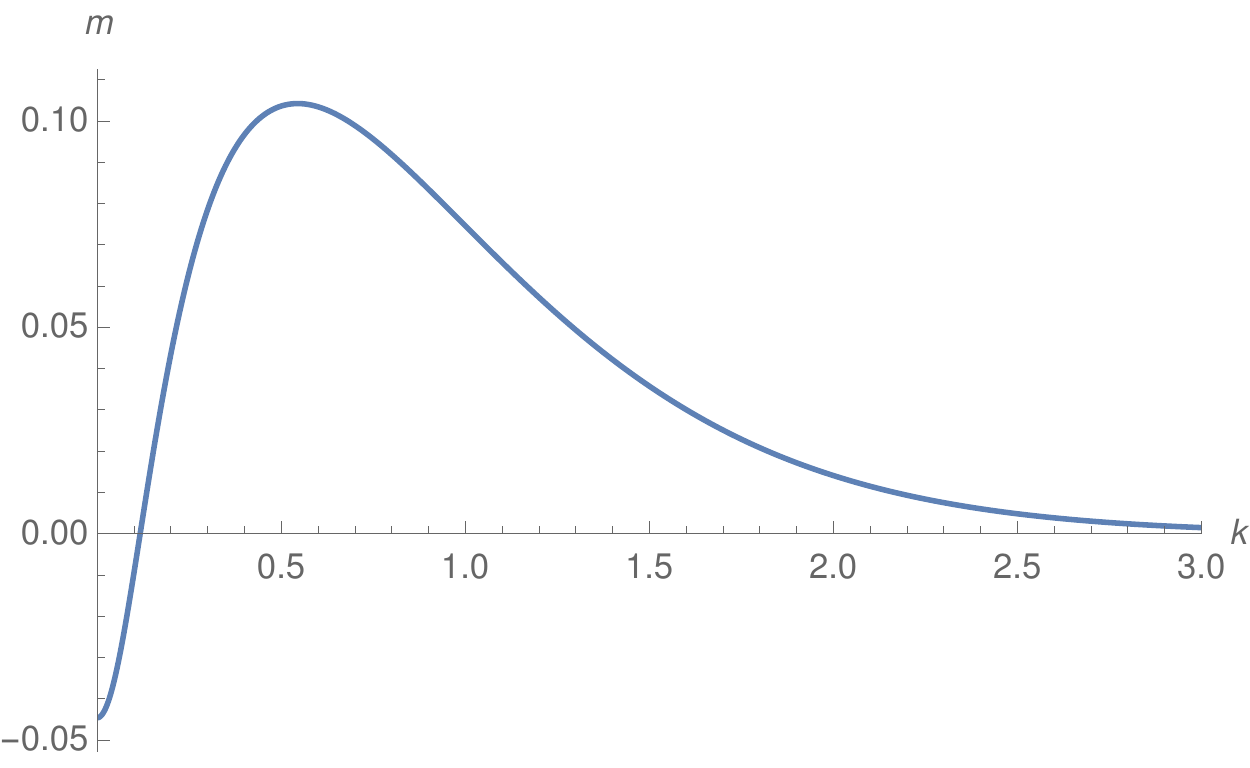}
\includegraphics[width=0.66\columnwidth]{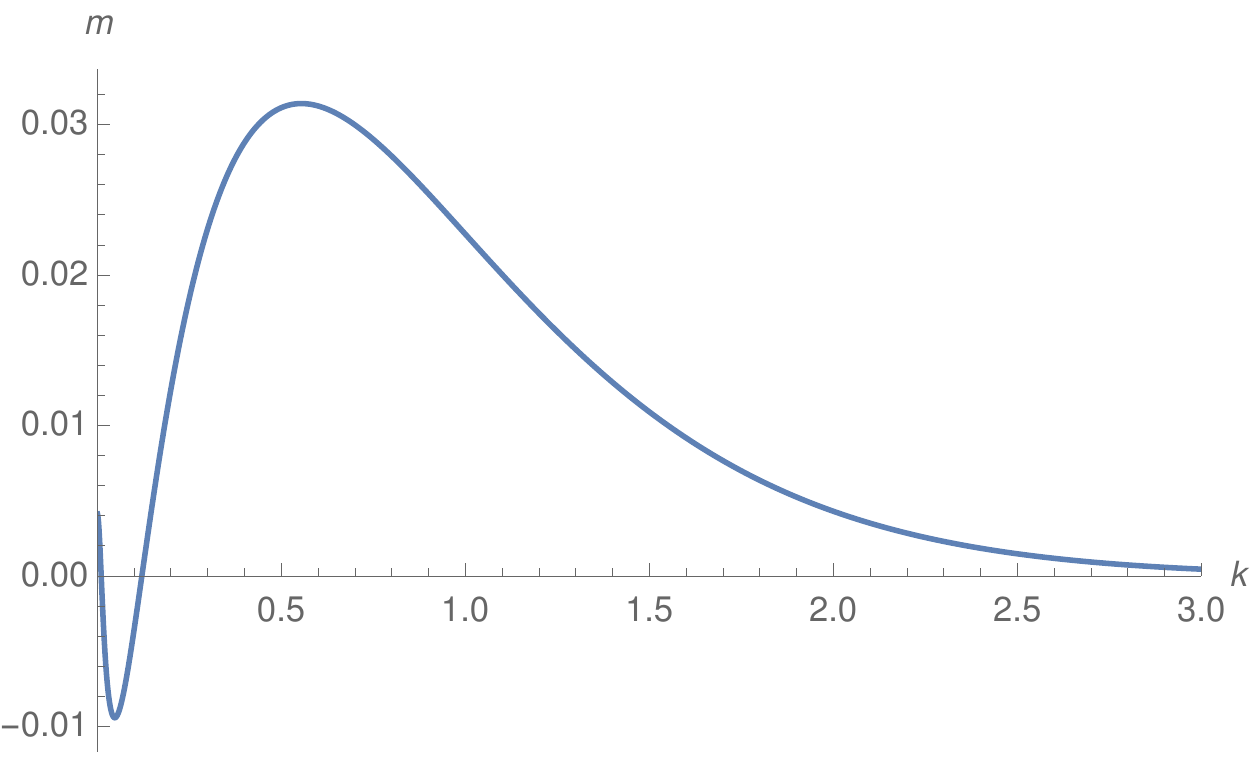}
\caption{
The constituent quark masses $m_c(k)$ in the chiral limit $m_0=0$, solutions of the mass gap equation,
from left to right for the ground state vacuum and first two replicas. 
}\label{fig:masssolutionII}
\end{figure}

\begin{figure}[t!]
\includegraphics[width=0.66\columnwidth]{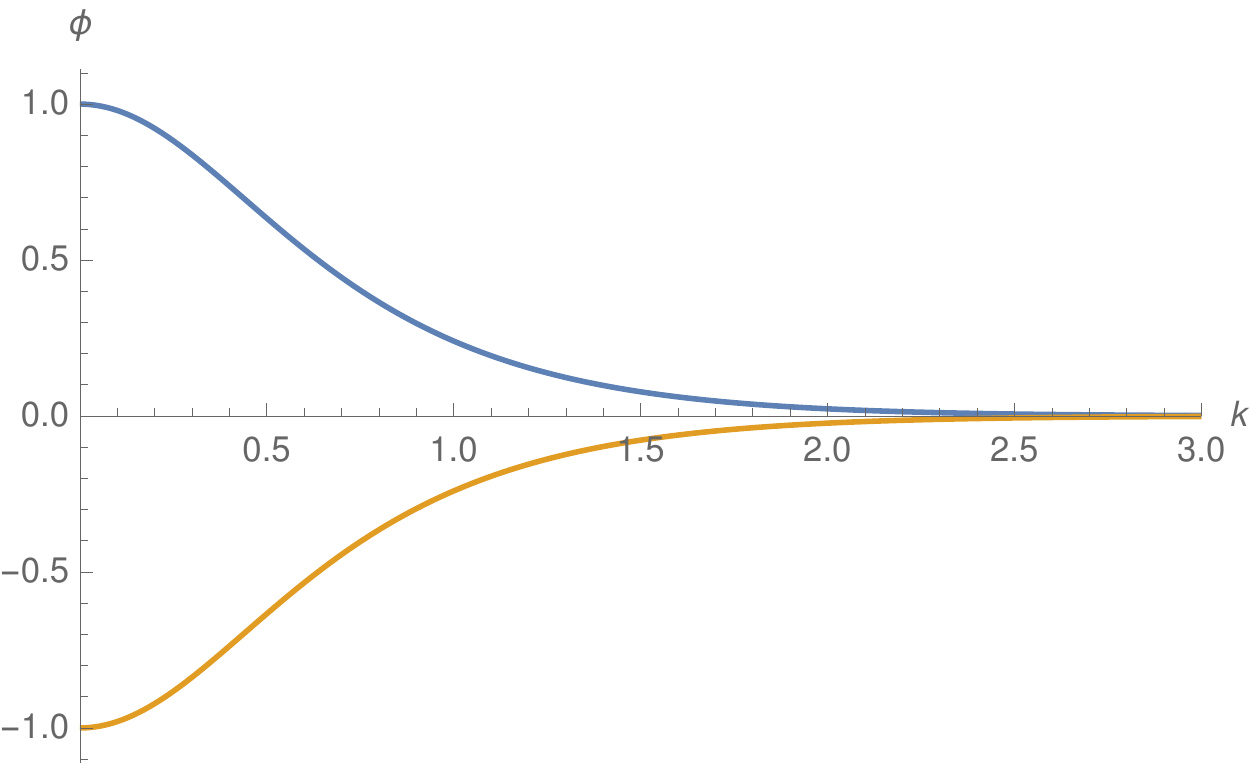} \hspace{1pt}
\includegraphics[width=0.66\columnwidth]{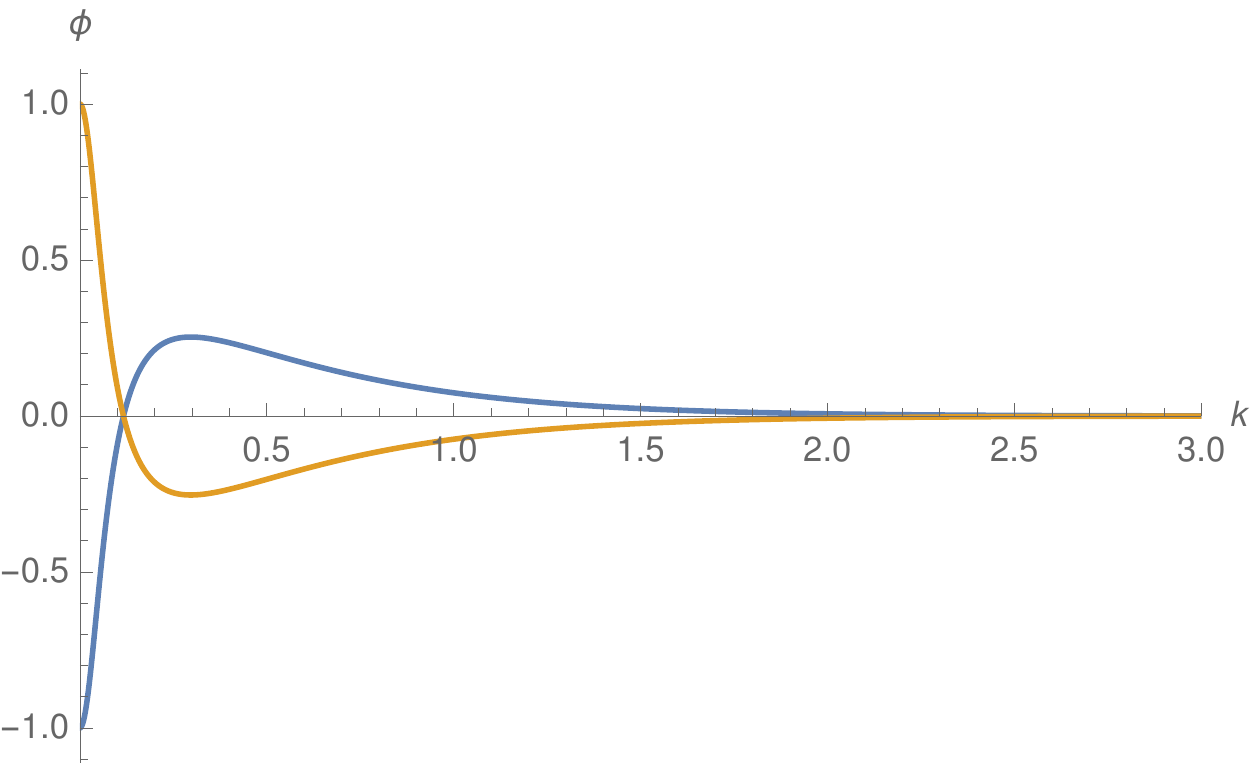} \hspace{1pt}
\includegraphics[width=0.66\columnwidth]{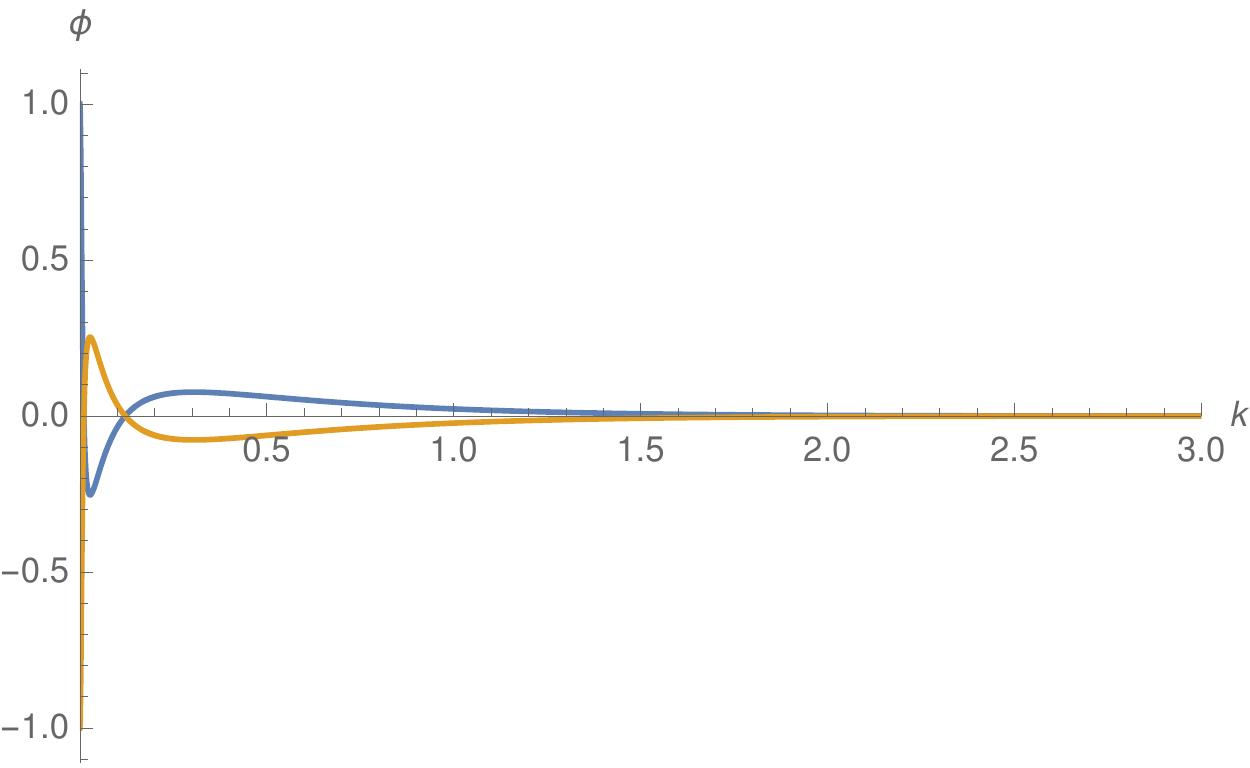}
\caption{The $J^{PC}=0^{-+}$,  $^1 S_0$ normalized pseudoscalar (P) radial wave functions  $\phi^+$ (in blue) and $\phi^-$ (in yellow), from left to right for the ground state vacuum and first two replicas, in dimensionless in units of $K_0=1$. Because the normalization diverges in the chiral limit, we arbitrarily normalize the wave functions with $\phi(0)=1$. Importantly, the wavefunctions $\phi^\pm$ are identical to $\pm \sin \varphi(k) =\pm m_c(k) /\sqrt{ k+ m_c(k)}$ as is easily verified from Fig.(\ref{fig:masssolution}).}
\label{fig:pseudoscalar}
\end{figure}

%
%
\newpage
\section{Numerical solution of the mass gap and boundstate equations.}

\begin{figure*}[t!]
\includegraphics[width=0.66\columnwidth]{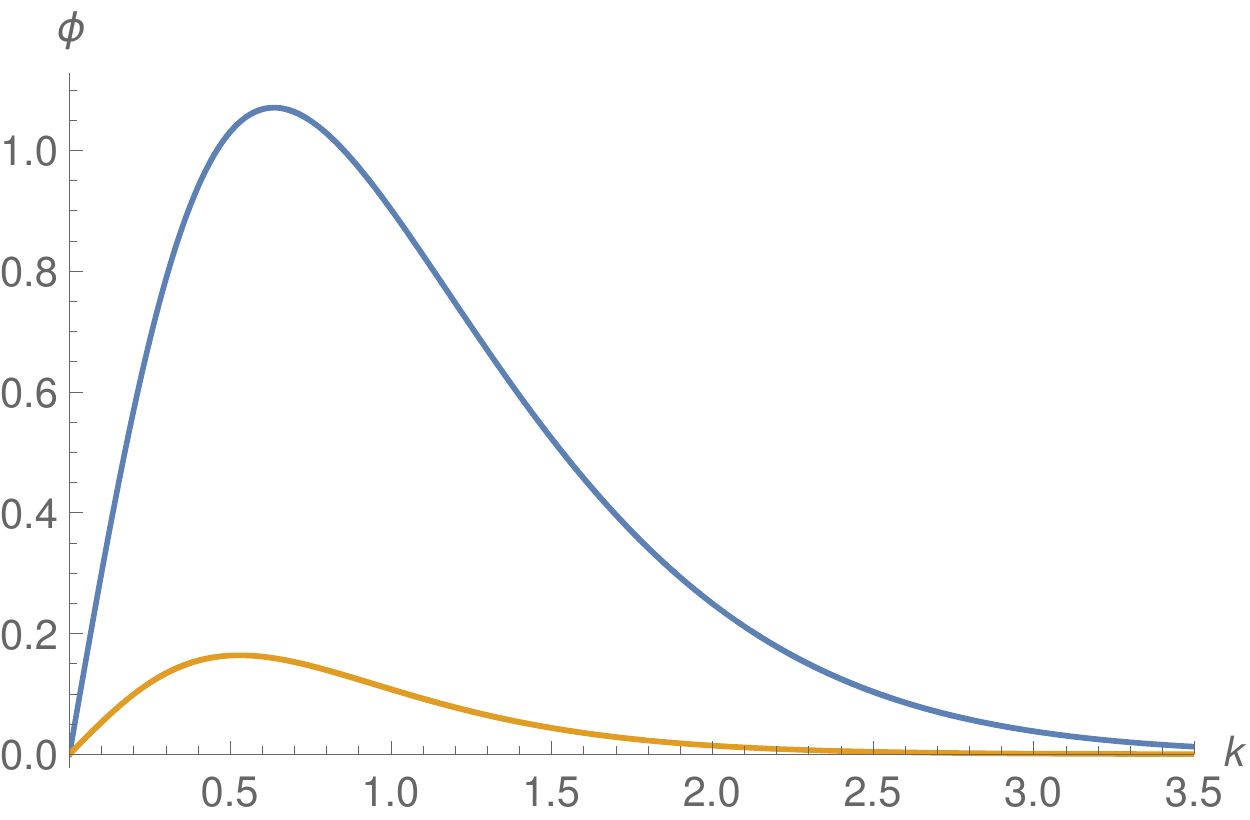} \hspace{1pt}
\includegraphics[width=0.66\columnwidth]{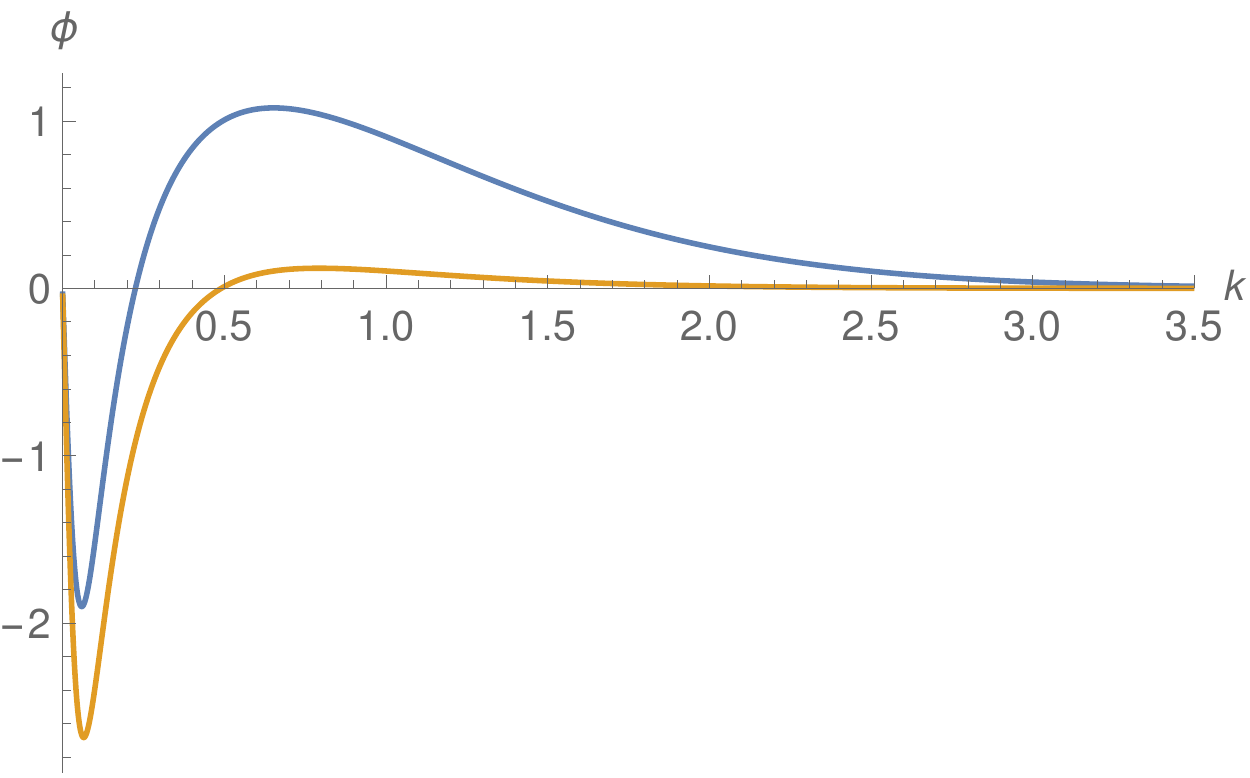} \hspace{1pt}
\includegraphics[width=0.66\columnwidth]{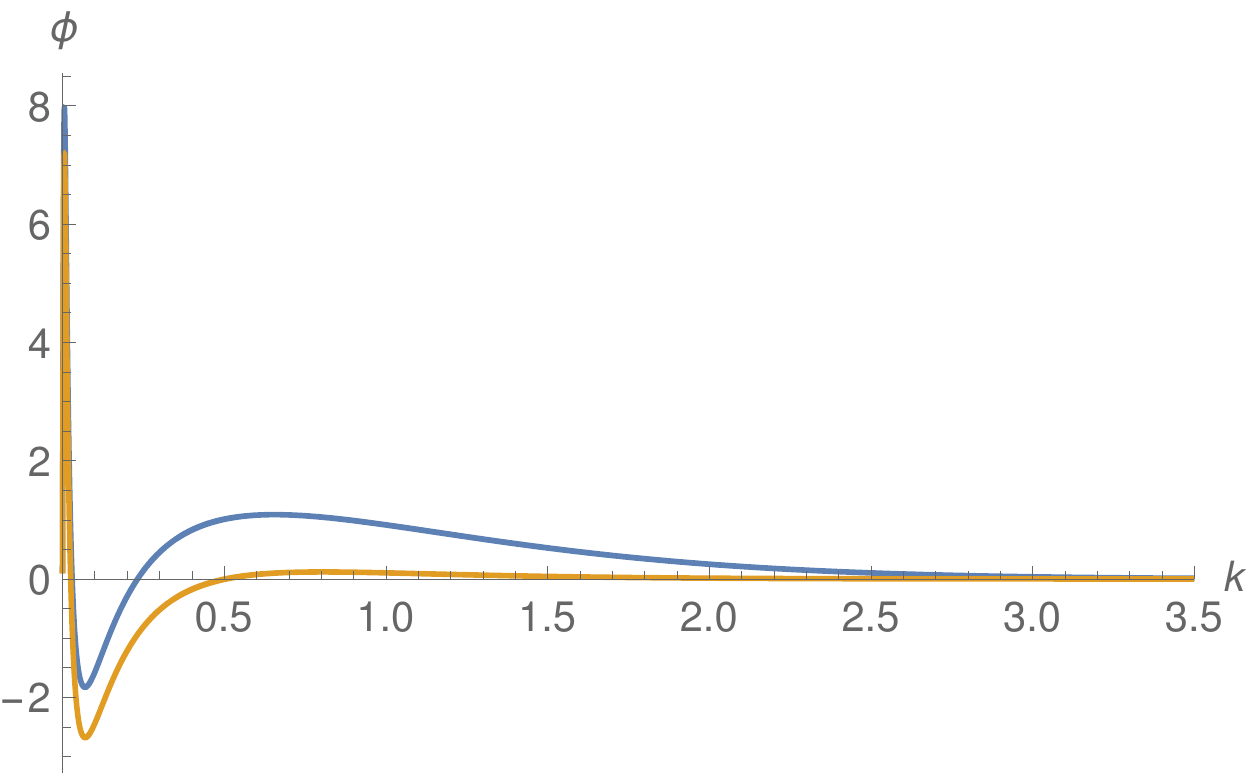}
\caption{The $J^{PC}=0^{++}$, $^3 P_0$ normalized scalar (S) radial wave functions $\phi^+$ (in blue) and $\phi^-$ (in yellow), from left to right for the ground state vacuum and first two replicas, in dimensionless in units of $K_0=1$. 
}\label{fig:scalar}
\end{figure*}
\begin{figure*}[t!]
\includegraphics[width=0.66\columnwidth]{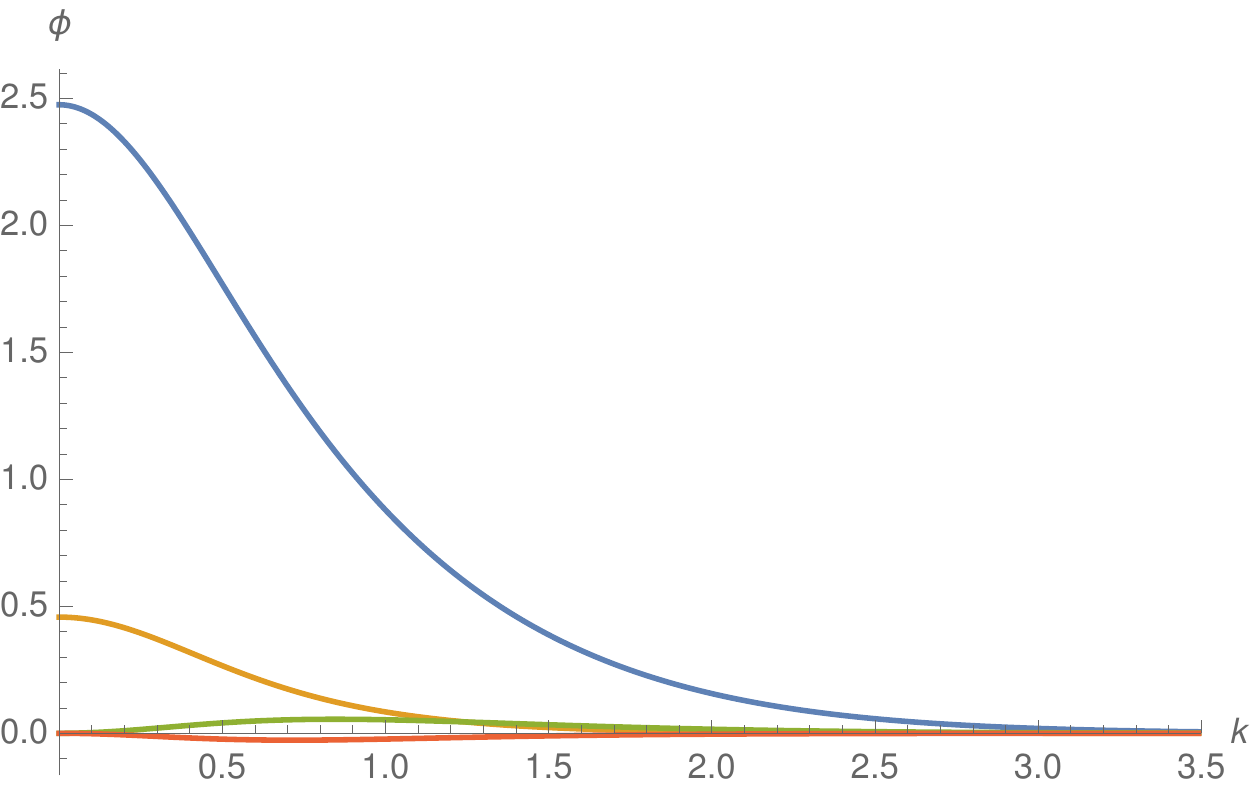} \hspace{1pt}
\includegraphics[width=0.66\columnwidth]{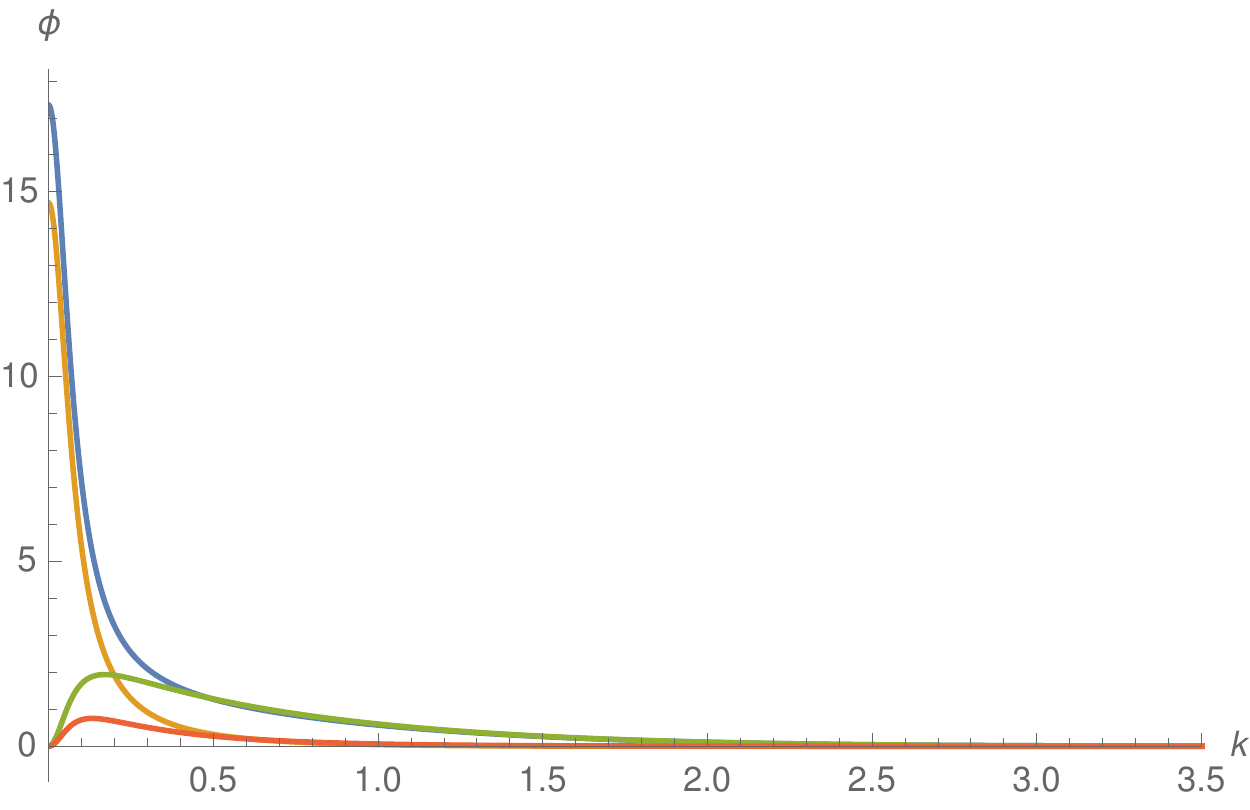} \hspace{1pt}
\includegraphics[width=0.66\columnwidth]{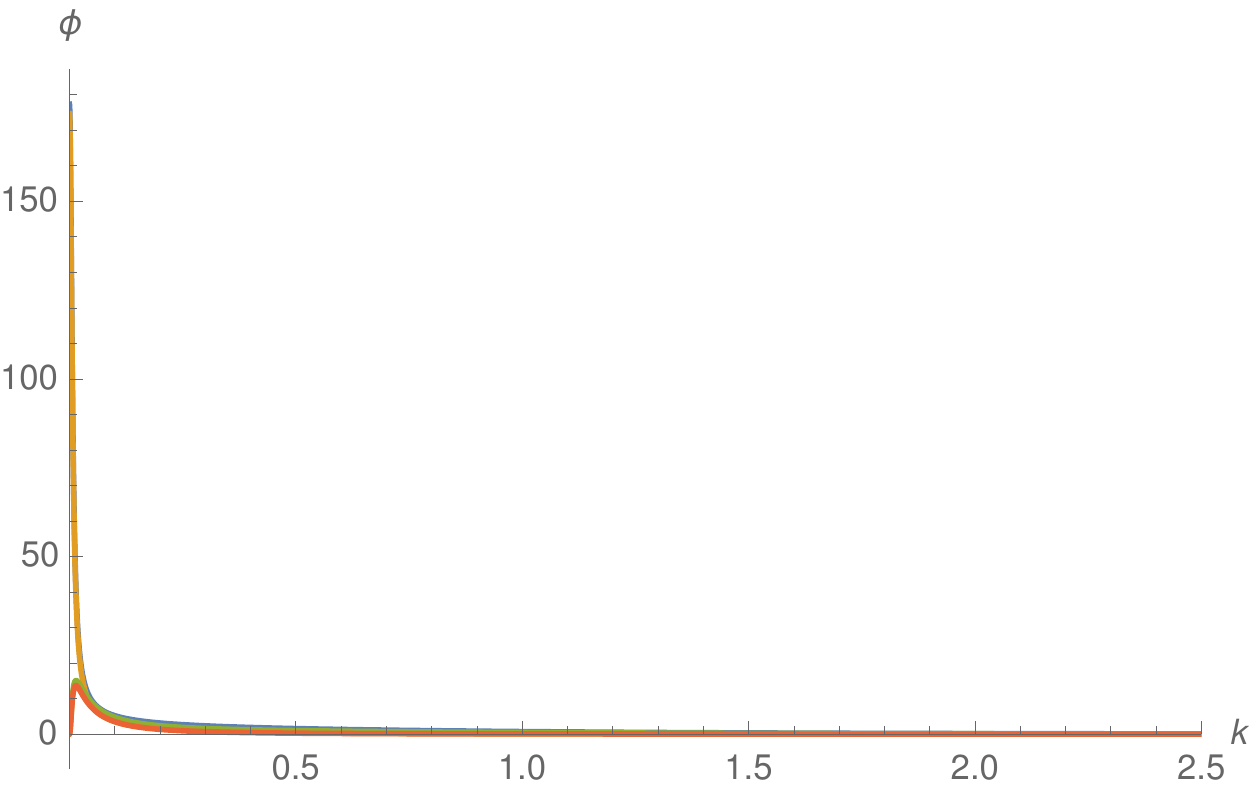}
\caption{The $J^{PC}=1^{--}$, $^3 P_1$ normalized vector (V) radial wave functions $\phi_0^+$ (in blue), $\phi_0^-$ (in yellow), $\phi_2^+$ (in green) and $\phi_2^-$ (in red) from left to right for the ground state vacuum and first two replicas,  in dimensionless in units of $K_0=1$. 
}\label{fig:vector}
\end{figure*}

\begin{figure*}[t!]
\includegraphics[width=0.66\columnwidth]{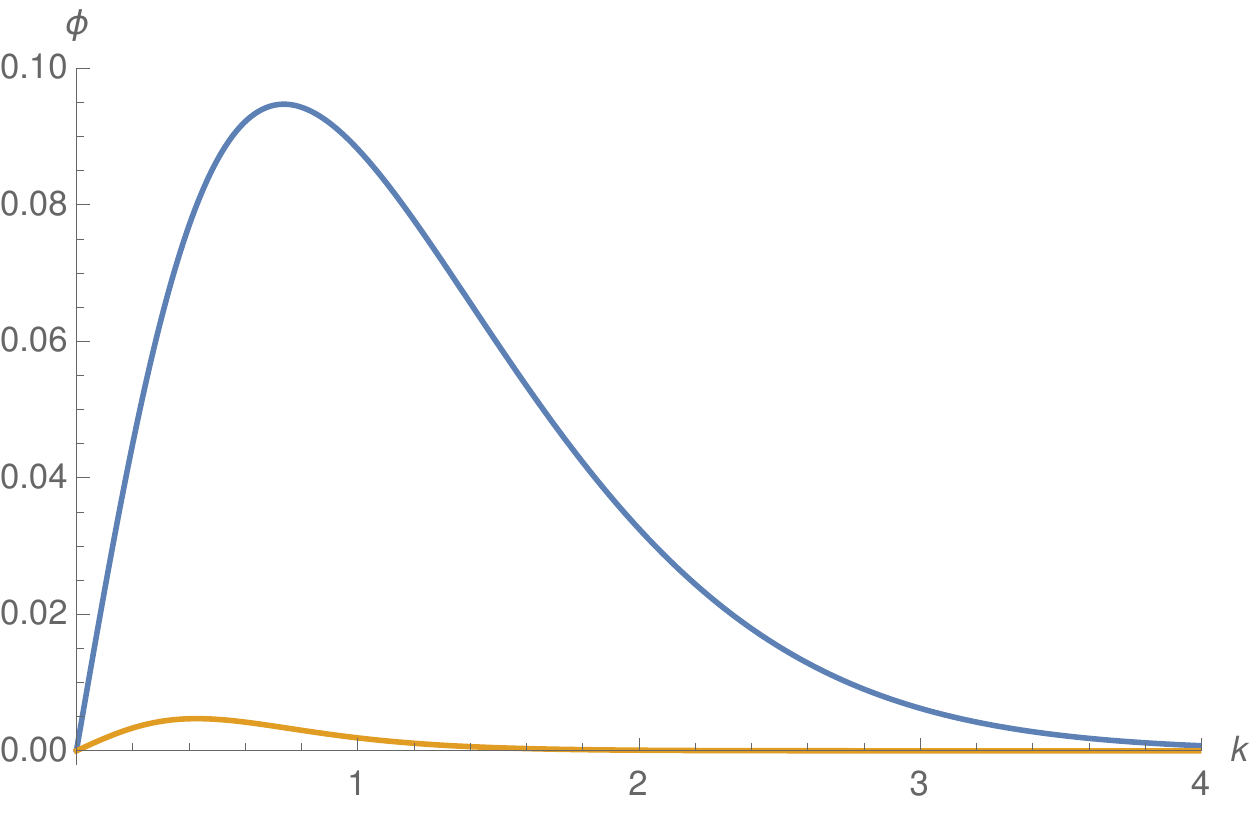} \hspace{1pt}
\includegraphics[width=0.66\columnwidth]{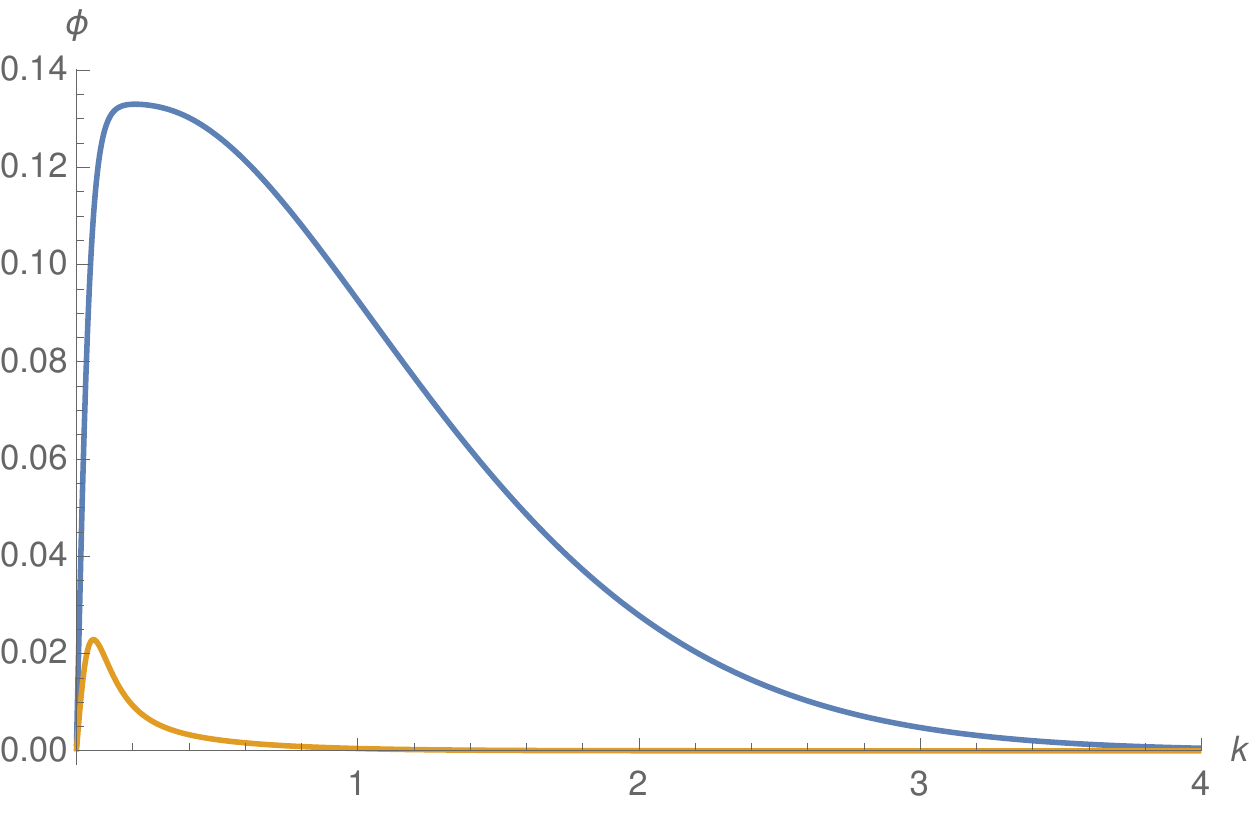} \hspace{1pt}
\includegraphics[width=0.66\columnwidth]{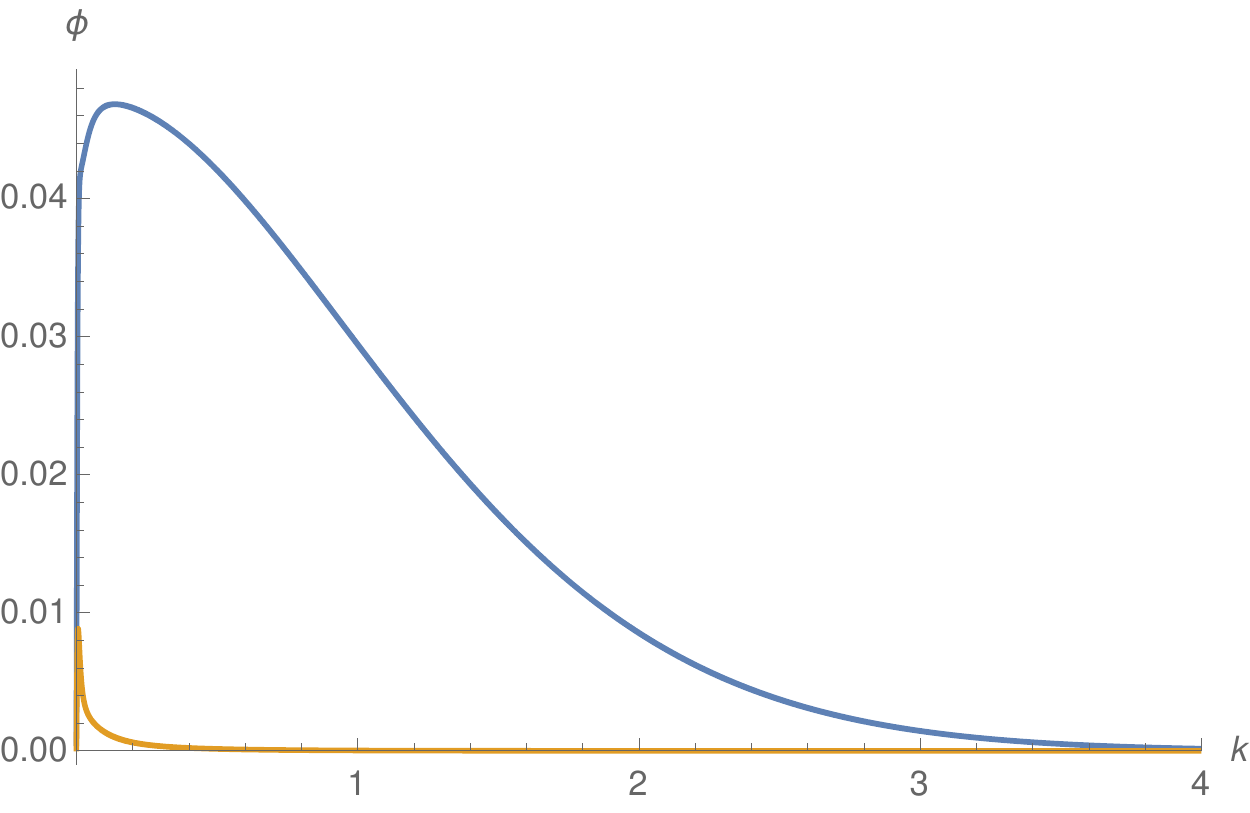}
\caption{The $J^{PC}=1^{+-}$, $^1 P_1$ normalized axial vector (A) radial wave functions $\phi^+$ (in blue) and $\phi^-$ (in yellow), from left to right for the ground state vacuum and first two replicas, in dimensionless in units of $K_0=1$. 
}\label{fig:axialvector1}
\end{figure*}

\begin{figure*}[t!]
\includegraphics[width=0.66\columnwidth]{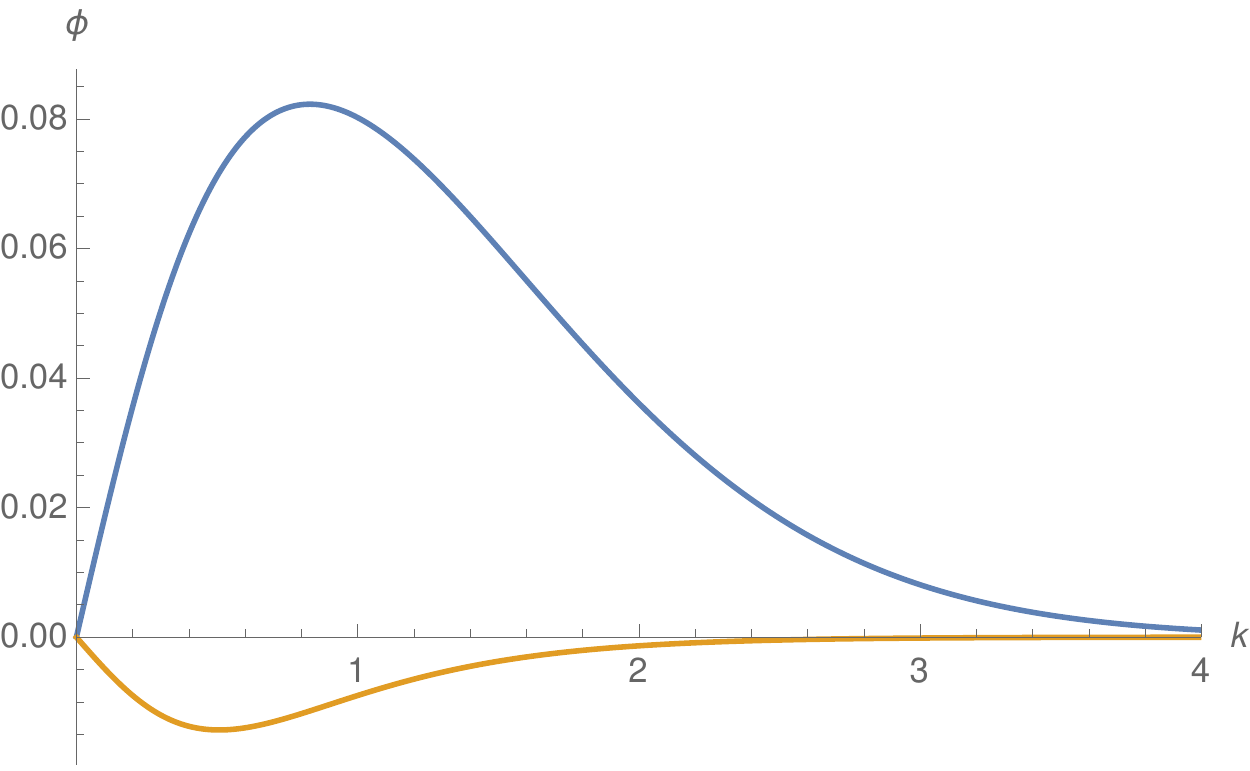} \hspace{1pt}
\includegraphics[width=0.66\columnwidth]{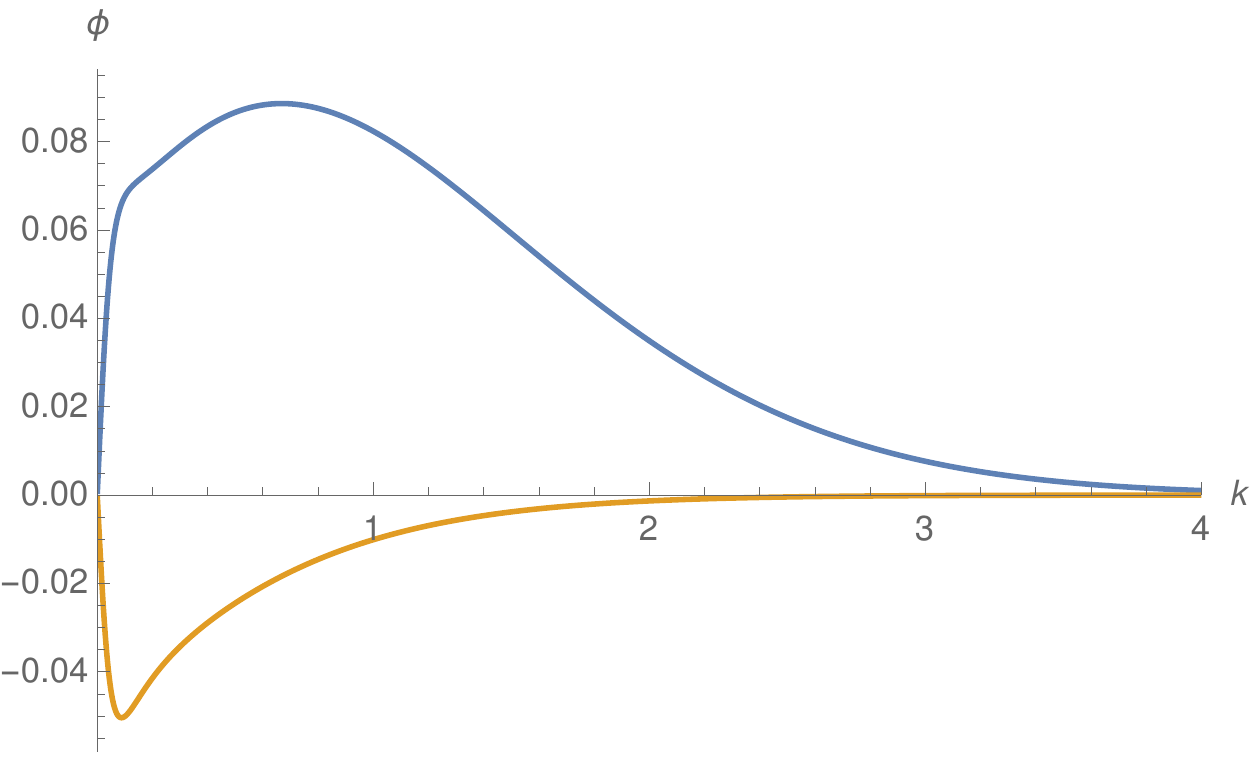} \hspace{1pt}
\includegraphics[width=0.66\columnwidth]{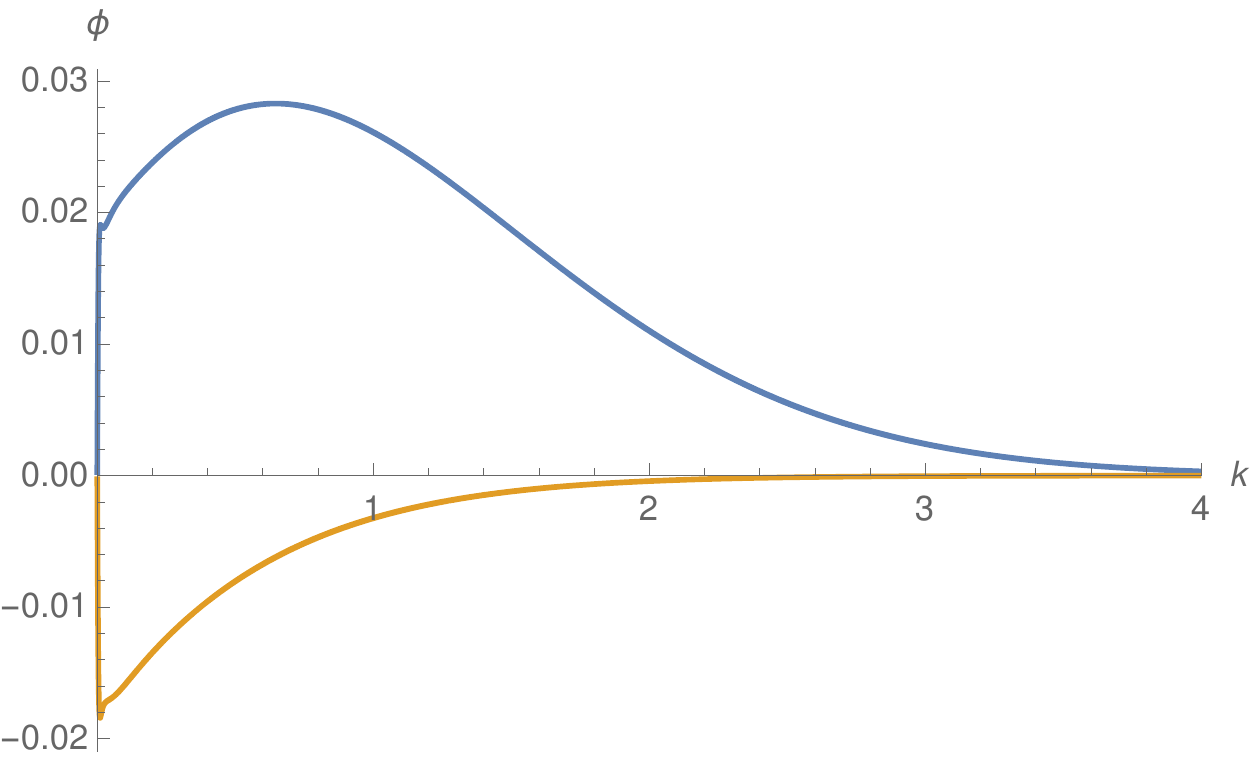}
\caption{The $J^{PC}=1^{++}$, $^3 P_1$ normalized axial vector (A) radial wave functions $\phi^+$ (in blue) and $\phi^-$ (in yellow), from left to right for the ground state vacuum and first two replicas,  in dimensionless in units of $K_0=1$. 
}\label{fig:axialvector2}
\end{figure*}

In each of the Eqs.(\ref{pseudoscalar} - \ref{axialvector}), the eigenvalue ${\cal M}$ must be understood as the excess mass of the quoted hadron above the masses of the corresponding supporting replica, namely, ${\cal R}i$. The true hadron masses are thus given by the sum of these two parcels.  

For the mass gap equation, we have the boundary conditions $m'(0)=0$ and $m(\infty)=0$. To solve numerically our differential equations, we use finite centred differences for the Laplacian. Since our mass gap equation is non-linear, we utilize the shooting method, so as to have the mass to vanish at a large enough UV momentum cutoff $K$ . We use a very fine mesh in momentum space because the replicas have nodes, quite close to the momentum origin. 

In Fig.(\ref{fig:masssolution}) we show the constituent running quark masses $m_c(k)$, corresponding both to the vacuum and to the first two replicas. 
Notice that in each replica, the running mass, $m(k)$, acquires an extra node.
The masses not only decrease when we go to a higher replica, but also the extra node amounts to a finer structure close to the momentum origin $k=0$.

We address now the main goal of this study: to show that the quark-antiquark bound states -- the mesons in the replicas -- all have real excess masses ${\cal M}$ -- see Eqs.(\ref{pseudoscalar} -- \ref{axialvector}). Our equation Salpeter bound state equations can be understood as an extension of the Shr\"odinger equation, since we now have positive $\phi^+$ and negative $\phi^-$ energy components of the wave functions, and this doubles the number of components. Nevertheless we have the same number 
of meson states as in the spectrum of the normal quark model. The mass splittings can also be related, as usual, to spin-tensor potentials.

For the bound state Salpeter equation, we also have similar boundary equations  to the mass gap equation: $u'(0)=0$ and $u(\infty)=0$ for the radial component in momentum space $u(k)=k \phi(k)$. We again use finite differences, and our linear differential equations are transformed into matrix equations. 

We use sparse matrices and matrix eigenvalues to compute the meson spectrum. Our results are obtained utilizing up to 100000 points in the discretization of the momentum, to comply both with the short distance nodes and with the large  vanishing distance for the amplitudes. In the case of the vector and axial vector potentials, this amounts to using very large matrices with size 400000 x 400000. We verify that our numerical results are stable for changes of both the number $N_k$  points in momentum space and momentum UV cutoff $K$.

With the effective quark masses $m(k)$, we can compute the chiral angle, using ${ m(k) / k}=\tan[ \varphi (k) ]$.
According to the chiral theorem on the pion Salpeter amplitude \cite{Bicudo:1989sh,Bicudo:2003fp}, the sine of the chiral angle $\sin[\varphi(k)]=m_c(k) /\sqrt{ k+ m_c(k)}$ should be proportional to the wave functions $\phi^{\pm}(k)$ of the pion. This is indeed clearly the case, when we compare the mass represented in Fig.(\ref{fig:masssolution}) with the wavefunction shown in Fig.(\ref{fig:pseudoscalar}). For the ground state vacuum and for the first two replicas, in Figs.(\ref{fig:pseudoscalar} and \ref{fig:scalar}), we show, respectively and in dimensionless in units of $K_0=1$, the wave functions $\phi^+$ and $\phi^-$, for the pseudoscalar meson and  the radial wave functions $\nu^+$ and $\nu^-$ for the scalar meson.  We notice that the number of nodes of these pseudoscalar wave functions depends on the replica where they are sitting, they have the same number of nodes as the constituent quark mass. This is expected due to the chiral theorem relating the pion wave function to the constituent quark mass \cite{Bicudo:1989sh,Bicudo:2003fp} .

We also show, in Figs.(\ref{fig:scalar} -- \ref{fig:axialvector2}), the radial wave functions of the scalar, vector and axial vector mesons. Notice how the wave functions change from one replica to another.
It is interesting to remark that, for the replicas, the excess masses of the mesons depend very little on the replica (and the physical vacuum) they are sitting in.

%
%
\section{Conclusion}

%
%
\begin{table}[t!] 
\begin{ruledtabular}
\begin{tabular}{c|c|c|c}
meson & in vacuum				& in replica 1$^1$ & in replica 2$^2$	\\ \hline
P \ $(J^{PC}=0^{-+})$ & 0.00 & 0.00 & 0.00 \\
P* $(J^{PC}=0^{-+})$ & 5.539 & 5.577 &  5.581 \\
S \ $(J^{PC}=0^{+ +})$ & 3.266 & 3.253 &  3.247 \\
V$_0$ $(J^{PC}=1^{--})$  & 2.686 & 2.823 & 2.836 \\
V$_2$ $(J^{PC}=1^{--})$ & 4.965 & 4.635 & 4.599 \\
A  \ $(J^{PC}=1^{+ -})$ & 4.103 & 3.784  & 3.723\\
A  \ $(J^{PC}=1^{+ +})$ & 4.665 & 4.602 & 4.596 \\
\end{tabular}
\end{ruledtabular}
\caption{
Masses of the mesons in the ground state vacuum (the true one) and excess masses$^{1,2}$ for the first two replicas, in units of $K_0$. $^{(1)}$ Setting $K_0=300 MeV$, for ${\cal R}1$ and for a bubble radius of 5 fm, we get a mass of 45 GeV, the mass of ${\cal R}1$ that must be added to the quoted masses in the table.  $^{(2)}$ For ${\cal R}2$ we must add 49 GeV). For the mesons we show the pseudoscalar, the first excitation of the pseudoscalar, the scalar, the vector (mostly s-wave), the first excitation of the vector (mostly d-wave) and the two different axial vectors.
}
\label{tab:spectrum}
\end{table}

The real nature of the excess masses of the mesons, constitute the main result of this work. 

Prior to this study, we were not sure what the masses of the mesons would be in the excited replica vacua. Since tachyons have been shown to exist in the false - chiral invariant - vacuum, we could possibly have had tachyons n the replicas. The tachyonic free nature of the replicas, with the excess masses having real values, show that there is no tachyonic-driven instability of the replicas and thus they are metastable.

Our final results are shown in Table \ref{tab:spectrum}.  

As an outlook, it would be interesting to extend the present study to study the effects of a more realistic potential, such as the linear confining potential with $\alpha=1$ (see Fig.(\ref{fig:replicaslinear})) or of a small finite current mass $m_0$ (see Fig.(\ref{fig:masssolution})). We leave these interesting, but cumbersome, case studies for future investigations.

As it stands, we cannot have all the low energy properties of hadronic
physics due to $S\chi SB$ without having replicated states as a
subproduct,  Although probably quite demanding experimentally, it would
nevertheless  be extremely interesting to look wether  such  metastable
replicas  do actually exist in full QCD.

\begin{acknowledgements}

P.B.\ and J.E.R.\ acknowledge the support of CeFEMA\mbox{} under the FCT contract for R\&D Units UID/CTM/04540/2013, and P.B.\ acknowledges the FCT grant CERN/FIS-COM/0029/2017.

\end{acknowledgements}


\bibliographystyle{apsrev4-1}
\bibliography{Bibliografia_1}

\end{document}